\documentclass[floatfix,showpacs,showkeys,11pt]{revtex4}
\usepackage{epsfig}
\usepackage{epsf}
\usepackage{amsmath}
\usepackage{amssymb}
\begin{document}

\title{Baryon spectra and non-strange baryon strong decays in the chiral
SU(3) quark model}

\author{H.M. Zhao$^{1}$}
\author{P.N. Shen$^{5,1,4}$}
\author{Y.B. Ding$^{2}$}
\author{X.Q. Li$^{3,4}$}
\author{B.S. Zou$^{1,4,5}$}
\affiliation{$^1$Institute of High Energy Physics, Chinese Academy
of Sciences,
P.O.Box 918(4), Beijing 100049, China \\
$^2$Department of Physics, Graduate University of Chinese Academy
of Sciences, Beijing 100049, China\\
$^3$Department of Physics, Nankai University, Tianjin 300071, China\\
$^4$Institute of Theoretical Physics, Chinese Academy of Sciences, P.O.Box 2735, China\\
$^5$Center of Theoretical Nuclear Physics, National Laboratory of
Heavy Ion Accelerator, Lanzhou 730000, China}

\date{\today}

\begin{abstract}
In the framework chiral SU(3) quark model, the baryon spectra within
the band of $N\leq 2$ are studied, and the effect of the confining
potential in different configurations, namely the $\Delta$-mode and
Y-mode are compared. In the same way, the baryon spectra in the
extended chiral SU(3) quark model, in which additional vector meson
exchanges are introduced, are also calculated. It is shown that a
reasonable baryon spectrum in the chiral SU(3) quark model can be
achieved no matter whether the $\Delta$-mode or the Y-mode confining
potential is employed. In the extended chiral SU(3) quark model,
several energy levels are further improved. The resultant binding
energies of excited baryon states in different confining modes
deviate just by a few to several tens MeV, and it is hard to justify
which confining mode is the dominant one. The non-strange baryon
strong decay widths are further discussed in the point-like meson
emission model by using the wave-function obtained in the spectrum
calculation. The resultant widths can generally explain the
experimental data but still cannot distinguish which confining mode
is more important in this simple decay mode.
\end{abstract}



\maketitle

\section{Introduction}
In the framework of the chiral SU(3) quark model \cite{ZhangZY1997},
which is extropolated from the SU(2) linear $\sigma$ model
\cite{Fernandez}, a unified description of the experimental data on
the masses of the baryon ground states, the binding energy of
deuteron, and the baryon-baryon scattering has successfully been
achieved \cite{ZhangZY1997,DaiLR03,HuangF03}. Later, this model has
been applied to the study of multi-quark systems to predict new
dibaryons and explain newly observed hadron states
\cite{LiQBNP01,HuangFPLB04}. Whether this model can also describe
the baryon spectrum in a reasonable extent would be one more place
to confirm the reliability of the model in studying the hadron
structure and the hadron-hadron scattering in the quark degrees of
freedom.

In the chiral SU(3) quark model, the short range perturbative
effect of QCD is generally characterized by the one-gluon-exchange
(OGE) potential, the medium range non-perturbative effect of QCD
is mainly described by one-Goldstone-bosons exchange (OBE)
potentials, and the long distance non-perturbative effect is
commonly depicted by a phenomenal confining potential, say a
harmonic confinement potential.

In terms of the OGE quark model, we calculated the baryons spectra
with different confining potential modes \cite{commu}, and such
potentials were derived from the flux tube model
\cite{Carlson,Isgur85} and later were confirmed by lattice QCD
(LQCD) calculations \cite{Bali}. For the baryon system, there exist
two confining modes. In the first mode, called $\Delta$-mode, the
confining potential can approximately be described by a sum of
two-body confining potentials. The second mode associates with a
genuine three-body interaction, called Y-mode. In recent years, with
the development of the fast computer, more and more LQCD
calculations on three-quark potential have been carried out. Typical
works on this aspect are done by Takahashi {\it et
al.}\cite{Takahashi} and Alexandrou {\it et al.}\cite{Alexandrou}.
Takahashi {\it et al.} more accurately considered three quark Wilson
loops and advocated that the Y-shape confining mode is the dominant
confining mode in baryon. But, Alexandrou {\it et al.} believed that
the $\Delta$-mode is favored at least in the distances smaller than
$1.2fm$. Different from other works
\cite{Isgur86,Stancu,Kuzmenko,Narodetskii,Brambilla}, we
respectively calculated the contributions of these two confining
potential modes directly, and found that by employing either the
$\Delta$-mode or Y-mode confining potential, one can achieve
reasonable baryon spectra in OGE quark model \cite{commu}. In this
paper we evaluate the spectra and decay properties of baryons with
these two confining modes in the chiral SU(3) quark model, and hope
that which confining mode is more important in baryon can be
explored.

In the next section, the baryon spectrum in the chiral SU(3) quark
model is discussed and the effect of different confining modes is
compared. The baryon spectrum in the extended chiral SU(3) quark
model is further examined in section III, and in section IV the
decays of baryons are calculated in a point-like meson emission
model. Finally, the conclusion is drawn in section V.

\section{Baryon spectrum in chiral SU(3) quark model}

\subsection{Brief formulism}

In the chiral SU(3) quark model \cite{ZhangZY1997}, the interaction
induced by the chiral field describes the non-perturbative QCD
effect in the medium distance. The interacting Hamiltonian between
the quark and the chiral field can be written as
\begin{eqnarray}\label{hami}
H^{ch}_I = g_{ch} F({\bm q}^{2}) \bar{\psi} \left( \sum^{8}_{a=0}
\sigma_a \lambda_a + i \sum^{8}_{a=0} \pi_a \lambda_a \gamma_5
\right) \psi,
\end{eqnarray}
where $g_{ch}$ is the coupling constant between the quark and the
chiral-field, $\lambda_{0}$ is a unitary matrix, $\lambda_{1}, ...,
\lambda_{8}$ are the Gell-Mann matrix of the flavor SU(3) group,
$\sigma_{0},...,\sigma_{8}$ denote the scalar singlet and octet
fields and $\pi_{0},..,\pi_{8}$ represent the pseudoscalar singlet
and octet fields, respectively. $F({\bm q}^{2})$ is a form factor to
describe the chiral-field structure \cite{ito90,amk91} and, as
usual, is taken as
\begin{eqnarray}\label{faca}
F({\bm q}^{2}) = \left(\frac{\Lambda^2}{\Lambda^2+{\bm
q}^2}\right)^{1/2},
\end{eqnarray}
where $\Lambda$ is the cutoff mass of the chiral field. It can be
verified that ${H}^{ch}_{I}$ is invariant under the infinitesimal
chiral SU(3) transformation.

With ${H}^{ch}_{I}$, the chiral-field-induced effective quark-quark
potentials can be written as:
\begin{eqnarray}
V^{ch}_{ij}=\sum^{8}_{a=0}V_{\sigma_a}(\bm
r_{ij})+\sum^{8}_{a=0}V_{\pi_a} (\bm r_{ij}), \label{eq:potentialch}
\end{eqnarray}
where
\begin{eqnarray}
V_{\sigma_a}({\bm r}_{ij})=-C(g_{ch},m_{\sigma_a},\Lambda)
X_1(m_{\sigma_a},\Lambda,r_{ij}) [\lambda_a(i)\lambda_a(j)] +
V_{\sigma_a}^{\bm {l \cdot s}}({\bm r}_{ij}),
\end{eqnarray}
\begin{eqnarray}
V_{\pi_a}({\bm r}_{ij})=C(g_{ch},m_{\pi_a},\Lambda)
\frac{m^2_{\pi_a}}{12m_{q_i}m_{q_j}} X_2(m_{\pi_a},\Lambda,r_{ij})
({\bm \sigma}_i\cdot{\bm \sigma}_j) [\lambda_a(i)\lambda_a(j)]
+V_{\pi_a}^{ten}({\bm r}_{ij}),
\end{eqnarray}
and
\begin{eqnarray}
V_{\sigma_a}^{\bm {l \cdot s}}({\bm r}_{ij})&=&
-C(g_{ch},m_{\sigma_a},\Lambda)\frac{m^2_{\sigma_a}}{4m_{q_i}m_{q_j}}
\left\{G(m_{\sigma_a}r_{ij})-\left(\frac{\Lambda}{m_{\sigma_a}}\right)^3
G(\Lambda r_{ij})\right\} \nonumber\\
&&\times[{\bm L \cdot ({\bm \sigma}_i+{\bm
\sigma}_j)}][\lambda_a(i)\lambda_a(j)],
\end{eqnarray}
\begin{eqnarray}
V_{\pi_a}^{ten}({\bm r}_{ij})&=&
C(g_{ch},m_{\pi_a},\Lambda)\frac{m^2_{\pi_a}}{12m_{q_i}m_{q_j}}
\left\{H(m_{\pi_a}r_{ij})-\left(\frac{\Lambda}{m_{\pi_a}}\right)^3
H(\Lambda r_{ij})\right\} \nonumber\\
&&\times\left[3({\bm \sigma}_i \cdot \hat{r}_{ij})({\bm \sigma}_j
\cdot \hat{r}_{ij})-{\bm \sigma}_i \cdot {\bm
\sigma}_j\right][\lambda_a(i)\lambda_a(j)],
\end{eqnarray}
with
\begin{eqnarray}\label{eq:C}
C(g_{ch},m,\Lambda)=\frac{g^2_{ch}}{4\pi}
\frac{\Lambda^2}{\Lambda^2-m^2} m,
\end{eqnarray}
\begin{eqnarray}
\label{x1mlr} X_1(m,\Lambda,r)=Y(mr)-\frac{\Lambda}{m} Y(\Lambda
r),
\end{eqnarray}
\begin{eqnarray}
X_2(m,\Lambda,r)=Y(mr)-\left(\frac{\Lambda}{m}\right)^3 Y(\Lambda
r),
\end{eqnarray}
\begin{eqnarray}
Y(x)=\frac{1}{x}e^{-x},
\end{eqnarray}
\begin{eqnarray}
G(x)=\frac{1}{x}\left(1+\frac{1}{x}\right)Y(x),
\end{eqnarray}
\begin{eqnarray}\label{eq:H}
H(x)=\left(1+\frac{3}{x}+\frac{3}{x^2}\right)Y(x),
\end{eqnarray}
and $m_{\sigma_a}$ and $m_{\pi_a}$ being the masses of the scalar
meson and the pseudoscalar meson, respectively.

The short-range interaction in the model is mainly governed by the
one-gluon-exchange interaction
\begin{eqnarray}
V^{OGE}_{ij}=\frac{1}{4}g_{i}g_{j}\left(\lambda^c_i\cdot\lambda^c_j\right)
\left\{\frac{1}{r_{ij}}-\frac{\pi}{2} \delta({\bm r}_{ij})
\left(\frac{1}{m^2_{q_i}}+\frac{1}{m^2_{q_j}}+\frac{4}{3}\frac{1}{m_{q_i}m_{q_j}}
({\bm \sigma}_i \cdot {\bm \sigma}_j)\right)\right\}+V_{OGE}^{\bm
l \cdot \bm s},
\end{eqnarray}
with
\begin{eqnarray}
V_{OGE}^{\bm l \cdot \bm
s}=-\frac{1}{16}g_ig_j\left(\lambda^c_i\cdot\lambda^c_j\right)
\frac{3}{m_{q_i}m_{q_j}}\frac{1}{r^3_{ij}}{\bm L \cdot ({\bm
\sigma}_i+{\bm \sigma}_j)}.
\end{eqnarray}
For the interaction in the long-distance range, confining potential
dominates the behavior of the system. In this paper, two kinds of
confining modes are considered. The $\Delta$-mode confining
potential of a three quark system can generally be written as:
\begin{equation}
 V^{conf}_{\triangle}=\frac{1}{2}b{\sum\limits_{i<j}}{r_{ij}}+C_{\triangle},
 \label{eq:deltashape}
\end{equation}
where $b$ denotes the string tension, $C_{\triangle}$ is an overall
constant. The Y-mode confining potential is related to the energy of
the flux tube connecting three valance quarks, and the total length
of the flux tube should take the minimal value for stability. The
general form of the Y-mode confining potential is written as
\cite{Isgur86}
\begin{equation}
V^{conf}_{Y}=b{\sum\limits_{i=1}^{3}}\mid{\bf{r_{i}}-\bf{r_{0}}}\mid
+C_{Y}, \label{eq:Yshape}
\end{equation}
where $C_{Y}$ is an overall constant, and $\bf{r_{0}}$ is the
coordinate of the junction point. The rule for finding the location
of the junction point is the following: If all the inner angles of
the triangle with three constituent quarks sitting at the apexes of
the triangle are smaller than $2\pi/3$, the junction point is
located inside the triangle and the angles spanned by two flux tubes
are $2\pi/3$. If one of the inner angles of the triangle would take
a value equal to or greater than $2\pi/3$, the junction point would
be located at that apex. Let the lengths of the three sides of the
triangle be a, b and c, respectively. $L_{min}$ then can be
expressed as \cite{Kogut75,Isgur86,Takahashi}
\begin{equation}
L_{min}=\left\{\begin{array}{cc}[\frac{1}{2}(a^{2}+b^{2}+c^{2})+\frac{\sqrt{3}}
{2}\sqrt{(a+b+c)(-a+b+c)(a-b+c)(a+b-c)}\ ]^{1/2}
\\
if\ all\ the\ inner\ angles\ are\ smaller\ than\ 2\pi/3,\\&\\
a+b+c-max(a,b,c)\\if\ one\ of\ the\ inner\ angles\ is\ not\ smaller\
than\ 2\pi/3.\end{array}\right.\label{eq:Lmin}
\end{equation}

For the baryon systems , the total Hamiltonian can be written as
\begin{eqnarray}
\label{hamiSU(3)}
H=\sum_{i=1}^{3}T_{i}-T_{G}+\sum_{i<j=1}^{3}V_{ij}+V^{conf}~,
\label{eq:HamiltonianTOT}
\end{eqnarray}
with
\begin{eqnarray}
V_{ij}= V^{OGE}_{ij} + V^{ch}_{ij}.
\end{eqnarray}
Where $T_i$ and $T_G$ are the kinetic energy operators of the i-th
quark and the center of mass, respectively. The kinetic energy
operator in the semi-relativistic form is
\begin{equation}
T_i={\sum\limits_{i=1}^3}\sqrt{p_{i}^{2}+m_{i}^{2}}.
\label{eq:Momentum}
\end{equation}

As a physical baryon $B$, it has a certain spin ($J,J_Z$) and a
certain parity ($P$), where $J$ and $J_Z$ represent the quantum
number of the total angular momentum ${\mathbf {J}}={\mathbf
{L}}+{\mathbf {S}}$ and its magnetic quantum number, respectively,
and $P=(-1)^L$. Therefore, the wave function of the baryon $B$ with
quantum number $J^P$ should be constructed as
\begin{eqnarray}
|J,J_Z,P,B> &=& \Phi^{C} \sum_{M=-L}^{L}\sum_{S_{Z}}(J,J_{Z}\mid
L,M,S,S_{Z})\phi_B^{SF}(S,S_Z,\xi,\Sigma_B^{SF})\psi^N_{L,M}(\eta,\zeta,\Sigma^O),
\label{eq:eigenstates}
\end{eqnarray}
where $(J,J_{Z}\mid L,M,S,S_{Z})$ is the CG coefficient for L-S
coupling, $\Phi^C$ denotes the total color wave function which
should be totally antisymmetric,
$\phi_B^{SF}(S,S_Z,\xi,\Sigma_B^{SF})$ represents the $SU_{SF}(6)$
spin and flavor wave functions, $\psi^N_{L,M}(\eta,\zeta,\Sigma^O)$
is the spatial wave function with
$N=2(n_{\rho}+n_{\lambda})+l_{\rho}+l_{\lambda}$ being the principal
quantum number, $L$ and $M$ being the quantum number of the total
orbital angular momentum ${\mathbf {L}}={\mathbf
{l}}_{\rho}+{\mathbf {l}}_{\lambda}$ and its magnetic quantum
number, respectively, $\Sigma^O$ standing for the symmetry of the
spatial wave function, $\eta$ being the width parameter and $\zeta$
representing the aggregate of the spatial variables, and
$\phi_B^{SF}(S,S_Z,\xi,\Sigma_B^{SF})$ and
$\psi^N_{L,M}(\eta,\zeta,\Sigma^O)$ must be coupled to a symmetric
wave function. The detailed forms of these wave functions can be
found in Ref.~\cite{Flamm,close}.

The baryon spectrum is solved by using the variational method. The
matrix element of kinetic energy operator (\ref{eq:Momentum}) is
calculated in the momentum space \cite{Isgur86}. According to the
definitions of the coordinates
\begin{eqnarray}
\mathbf{R}&=&\frac{m_{1}{\mathbf{r_{1}}}+m_{2}{\mathbf{r_{2}}}+m_{3}{\mathbf{r_{3}}}}
{m_{1}+m_{2}+m_{3}}\nonumber\\
\bm{\rho}&=&\frac{1}{\sqrt{2}}(\mathbf{r_{1}}-\mathbf{r_{2}})\\
\bm{\lambda}&=&\sqrt{\frac{2}{3}}\left(\frac{m_{1}{\mathbf{r_{1}}}+m_{2}{\mathbf{r_{2}}}}
{m_{1}+m_{2}}-\mathbf{r_{3}}\right),\nonumber \label{eq:Jacobian}
\end{eqnarray}
the momenta in the rest frame are related to those in the center of
mass frame by
\begin{eqnarray}
\mathbf{p_{1}}&=&\frac{1}{\sqrt{2}}\mathbf{p}_{\rho}
+\frac{1}{\sqrt{6}}\mathbf{p}_{\lambda},\nonumber\\
\mathbf{p_{2}}&=&-\frac{1}{\sqrt{2}}\mathbf{p}_{\rho}
+\frac{1}{\sqrt{6}}\mathbf{p}_{\lambda},\\
\mathbf{p_{3}}&=&-\frac{2}{\sqrt{6}}\mathbf{p}_{\lambda}.\nonumber
\end{eqnarray}
The detailed evaluation of the matrix elements can be found in Ref.
\cite{commu} and references therein. In our numerical calculation,
the spin-orbital interaction is dropped out due to the weakness of
such an interaction showing in the experimental data of the baryon
spectrum.

\subsection{Determination of parameters}

There are four initial input parameters: the up (down) quark mass
$m_{u(d)}$, the strange quark mass $m_{s}$, the harmonic oscillator
frequency $\omega$ and the confining strength $b$. The up (down)
quark mass $m_{u(d)}$ and the strange quark mass $m_{s}$ are taken
to be commonly used values of $330 MeV$ and $470 MeV$, respectively.
$\omega$ is chosen to be 0.497 $GeV$, which will produce a radius of
about $0.5fm$ for a bare nucleon. The string tension is chosen,
according to the LQCD result \cite{Takahashi}, to be $0.20GeV^{-1}$
for the Y-mode and an additional factor of 0.53 for the
$\Delta$-mode. The other model parameters are fixed in the following
way. The chiral coupling constant $g_{ch}$ is fixed by
\begin{eqnarray}
\frac{g^{2}_{ch}}{4\pi} = \left( \frac{3}{5} \right)^{2}
\frac{g^{2}_{NN\pi}}{4\pi} \frac{m^{2}_{u}}{M^{2}_{N}},
\end{eqnarray}
with $g^{2}_{NN\pi}/4\pi=13.67$ being the empirical value. The
masses of exchanged mesons are adopted from the experimental data,
except for the $\sigma$ meson. In our calculation, the mass of
$\sigma$ is a free parameter and its value of 675 MeV is determined
by the best fit to the experimental data available
\cite{ZhangZY1997,HuangF03}. The cutoff momentum $\Lambda$ is taken
to be close to the chiral symmetry breaking scale
\cite{ito90,amk91}. After the parameters of chiral fields are fixed,
the one gluon exchange coupling constants $g_{u}$ and $g_{s}$ can be
determined by the mass splittings between $N$ and $\Delta$ and
between $\Sigma$ and $\Lambda$, respectively. The zero point
energies $C^{uu}$, $C^{us}$ and $C^{ss}$ in the $\Delta$-mode
confining potential are fixed by the masses of the ground state
baryons $N$, $\Lambda$ and the average masses of $\Xi$ and $\Omega$,
respectively. And the zero point energies $C^{N}$,
$C^{\Lambda}=C^{\Sigma}$, $C^{\Xi}$ and $C^{\Omega}$ in the Y-mode
confining potential case are fixed by the masses of $N$, $\Lambda$,
$\Xi$ and $\Omega$, respectively.

In the calculation, $\eta$ and $\eta'$ mesons are mixed by $\eta_1$
and $\eta_8$ with a mixing angle $\theta^{PS}$ of $-23^\circ$ as
usual. The mixing angle $\theta^{S}$ between $\sigma_0$ and
$\sigma_8$ is still an open issue because the structure of $\sigma$
meson is unclear and controversial. $35.264^\circ$ is adopted from
Ref. \cite{HuangF03} which indicates that $\sigma$ and $\epsilon$
are ideally mixed by $\sigma_0$ and $\sigma_8$.

The resultant model parameters are tabulated in Table
\ref{tab:parasu3}.

\begin{table}[htb]
\caption{\label{tab:parasu3} {\small Model parameters. The meson
masses and the cutoff masses are taken to be $m_{\sigma'}=980$ MeV,
$m_{\kappa}=980$ MeV, $m_{\epsilon}=980$ MeV, $m_{\pi}=138$ MeV,
$m_K=495$ MeV, $m_{\eta}=549$ MeV, $m_{\eta'}=957$ MeV,
$\Lambda=1100$ MeV.}}
\begin{center}
\begin{tabular}{ccccc}
 \hline\hline

 &~~~~~~~~~~~~~~~~~~~~~$\Delta$-shape &~~~~~~~~~~~~~~~~~~~~Y-shape \\
 \hline
$g_{u}$&~~~~~~~~~~~~~~~~~~~~0.900&~~~~~~~~~~~~~~~~~~~~0.900\\
$g_{s}$&~~~~~~~~~~~~~~~~~~~~0.955&~~~~~~~~~~~~~~~~~~~~0.955\\
$b_{\triangle}^{uu}(GeV/fm)$&~~~~~~~~~~~~~~~~~~~~ 1.01&~~~~~~~~~~~~~~~~~~~~--\\
$b_{\triangle}^{us}(GeV/fm)$&~~~~~~~~~~~~~~~~~~~~ 1.01&~~~~~~~~~~~~~~~~~~~~--\\
$b_{\triangle}^{ss}(GeV/fm)$&~~~~~~~~~~~~~~~~~~~~ 1.01&~~~~~~~~~~~~~~~~~~~~--\\
$C_{\triangle}^{uu}(MeV)$&~~~~~~~~~~~~~~~~~~~~ -1053 &~~~~~~~~~~~~~~~~~~~~--\\
$C_{\triangle}^{us}(MeV)$&~~~~~~~~~~~~~~~~~~~~ -936 &~~~~~~~~~~~~~~~~~~~~--\\
$C_{\triangle}^{ss}(MeV)$&~~~~~~~~~~~~~~~~~~~~ -751 &~~~~~~~~~~~~~~~~~~~~--\\
$b_{Y}^{N}(GeV/fm)$&~~~~~~~~~~~~~~~~~~~~--&~~~~~~~~~~~~~~~~~~~~ 0.91 \\
$b_{Y}^{\Lambda}(GeV/fm)$&~~~~~~~~~~~~~~~~~~~~--&~~~~~~~~~~~~~~~~~~~~ 1.01 \\
$b_{Y}^{\Xi}(GeV/fm)$&~~~~~~~~~~~~~~~~~~~~--&~~~~~~~~~~~~~~~~~~~~ 1.01 \\
$b_{Y}^{\Omega}(GeV/fm)$&~~~~~~~~~~~~~~~~~~~~--&~~~~~~~~~~~~~~~~~~~~ 1.01 \\
$C_{Y}^{N}(MeV)$&~~~~~~~~~~~~~~~~~~~~--&~~~~~~~~~~~~~~~~~~~~-1170\\
$C_{Y}^{\Lambda}(MeV)$&~~~~~~~~~~~~~~~~~~~~--&~~~~~~~~~~~~~~~~~~~~-1085\\
$C_{Y}^{\Xi}(MeV)$&~~~~~~~~~~~~~~~~~~~~--&~~~~~~~~~~~~~~~~~~~~-989\\
$C_{Y}^{\Omega}(MeV)$&~~~~~~~~~~~~~~~~~~~~--&~~~~~~~~~~~~~~~~~~~~-213\\
 \hline\hline
\end{tabular}
\end{center}
\end{table}

\subsection{Baryon spectrum}
\begin{figure}
\begin{center}\vspace*{0.5cm}
\parbox{.47\textwidth}{\epsfysize=6.5cm\epsffile{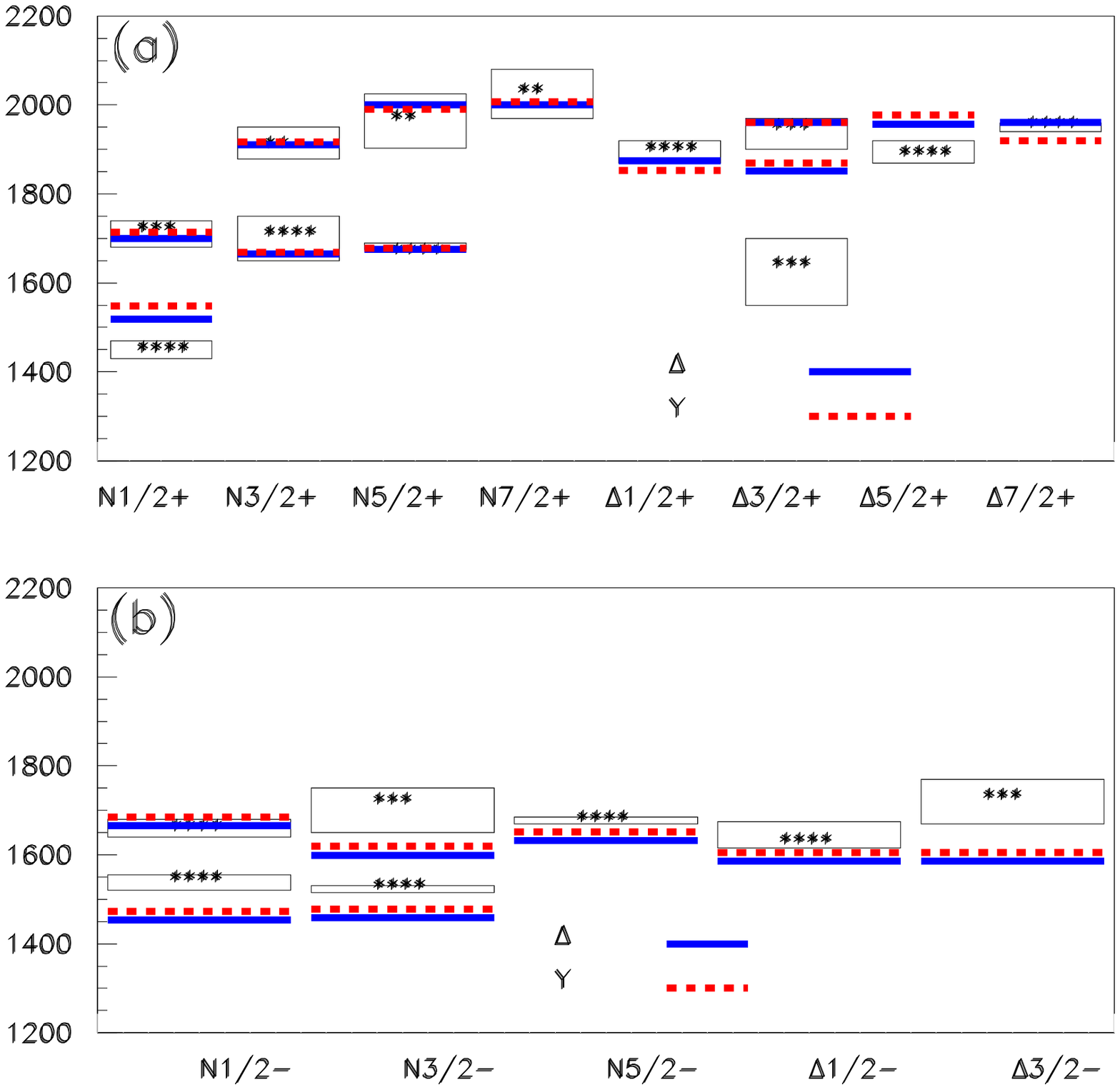}} \hfill
 \parbox{.47\textwidth}{\epsfysize=6.5cm \epsffile{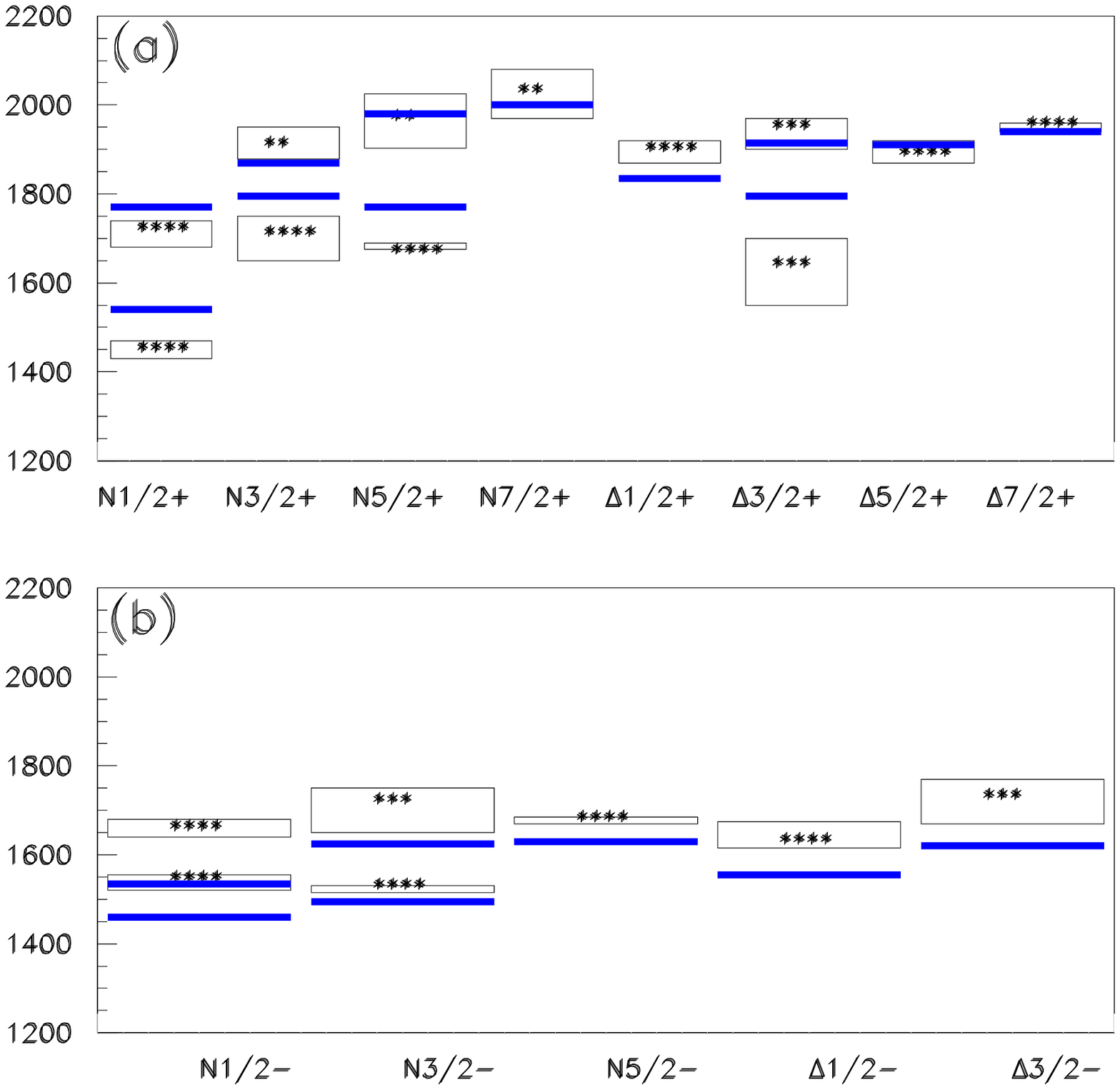}}\hfill
\vspace{0.5cm}
\parbox{.47\textwidth}{\caption{\label{fig:mixrenposneg}
{\small {The non-strange baryon in N $\leq$ 2 bands in the chiral
SU(3) quark model. Boxes show the experimental regions of the
resonances. The solid and dashed bars represent the results in the
$\Delta$-mode case and in the Y-mode case, respectively.}}}}\hfill
\vspace{-2.0cm}
\parbox{.49\textwidth}{\caption{\label{fig:isgurnposneg}
{\small {The non-strange baryon spectra in N $\leq$ 2 bands in
Ref.\cite{Isgur86}. Boxes show the experimental regions of the
resonances. The solid and dashed bars represent the results in the
$\Delta$-mode case and in the Y-mode case, respectively.}}}}
\end{center}
\vspace{1.5cm}
\end{figure}

In the chiral SU(3) quark model, the baryon spectra for $N\leq 2$
bands with $\Delta$-mode and Y-mode confining potentials are
calculated. The non-strange baryon spectra in N $\leq$ 2 bands are
plotted in Fig. \ref{fig:mixrenposneg} and also tabulated in Table
\ref{tab:mixrenposneg}.
\tabcolsep 0.1in \renewcommand\arraystretch{0.7}
\begin{table}
\begin{small}
\begin{center}
\caption{\label{tab:mixrenposneg}{\small {Non-strange baryon masses
in N $\leq$ 2 bands (in $MeV$). Experimental date are taken from
\cite{PDG}. }}} \vspace{0.3cm}
\begin{tabular}{cccccccc}
\hline\hline
State & $\Delta$-mode&Y-mode&experimental data\\
 \hline
$N^{*}\frac{1}{2}^{+}$& 939&939&939&****\\
 & 1519&1548&1430-1470&****\\
 & 1700&1714&1680-1740&***\\
 & 1970&1961&1885-2125&*\\
 & 2063&2056 &\\
\hline
$\Delta^{*}\frac{1}{2}^{+}$& 1874&1853&1870-1920&****\\
 & 1949&1942&\\
\hline
$N^{*}\frac{3}{2}^{+}$& 1665&1669&1650-1750&****\\
 & 1910&1917&1879-1951&**\\
 & 1977&1970&\\
 & 2051&2048&\\
 & 2357&2366\\
\hline
$\Delta^{*}\frac{3}{2}^{+}$& 1232&1232&1232&****\\
 & 1852&1870&1550-1700&***\\
 & 1961&1961&1900-1970&***\\
 & 2000&2024&\\
\hline
$N^{*}\frac{5}{2}^{+}$& 1675&1678&1675-1690&****\\
 & 2001&1990&1903-2025&**\\
 & 2356&2365\\
\hline
$\Delta^{*}\frac{5}{2}^{+}$& 1957&1977&1870-1920&****\\
 & 2028&2020&1724-2200&**\\
\hline
$N^{*}\frac{7}{2}^{+}$& 2000&2007&1970-2080&**\\
\hline
$\Delta^{*}\frac{7}{2}^{+}$& 1961&1919&1940-1960&****\\
 \hline\hline
$N^{*}\frac{1}{2}^{-}$& 1454&1473&1520-1555&****\\
 & 1665&1685&1640-1680&****\\
$\Delta^{*}\frac{1}{2}^{-}$& 1586&1605&1615-1675&****\\
\hline
$N^{*}\frac{3}{2}^{-}$& 1459&1478&1515-1530&****\\
& 1599&1619&1650-1750&***\\
$\Delta^{*}\frac{3}{2}^{-}$& 1586&1605& 1670-1770&***\\
\hline
$N^{*}\frac{5}{2}^{-}$& 1632&1652&1670-1685&****\\
\hline\hline
\end{tabular}
\end{center}
\end{small}
\end{table}
In this figure, the solid and dashed bars denote the results in the
$\Delta$-confining mode and in the Y-confining mode, respectively.
It is shown that no matter which confining mode is employed, most of
resultant resonances are fairly well located within experimental
errors. The differences of resultant resonances by using different
confining modes are rather small and most of them are in several
tens MeV to several MeV. For comparison, the results by using the
so-called relativistic quark model \cite{Isgur86}, where the
semi-relativistic corrections were considered and the OGE and
confining potentials were employed only, are also given in Fig.
\ref{fig:isgurnposneg}. Comparing with Fig. \ref{fig:isgurnposneg},
in our model, most resonances in positive-parity sector, except
$\Delta^{*}({\frac{3}{2}^{+}},1600)$ which is 200 MeV higher than
the experimental center value, can be fitted to the experimental
data much better. The reason for worse fitting of
$\Delta^{*}({\frac{3}{2}^{+}},1600)$ can be attributed to the fact
that the contribution of the contact term for this state is much
larger than that for the other states in this band (referring to the
hyperfine matrix elements in the table in Ref.\cite{Isgur7879}). To
avoid the singular behavior of the wave function at the origin due
to the $\delta({\mathbf {r}})$ function in the contact term in the
potential, the authors of Ref. \cite{Isgur86} employed a smearing
factor and also included the relativistic corrections for the terms
related to the quark mass. As a consequence, the mass of
$\Delta^{*}({\frac{3}{2}^{+}},1600)$ in their calculation is smaller
than ours but is still higher than the upper limit of experimental
error bar. For the negative parity sector, our results are similar
to Ref.\cite{Isgur86}, and our $N^{*}({\frac{1}{2}^{+}},1650)$ fits
the experimental data much better. Our resultant mass of the Roper
resonance ($N^{*}({\frac{1}{2}^{+}},1440)$) is still higher than
that of the first orbital resonance $N^{*}({\frac{1}{2}^{-}},1535)$,
which is similar to the predictions in Ref.\cite{Isgur86}. In recent
years, some one argued that the Roper resonance is not a pure baryon
\cite{Page02}.

\begin{figure}
\begin{center}\vspace*{0.5cm}
\parbox{.47\textwidth}{\epsfysize=6.5cm\epsffile{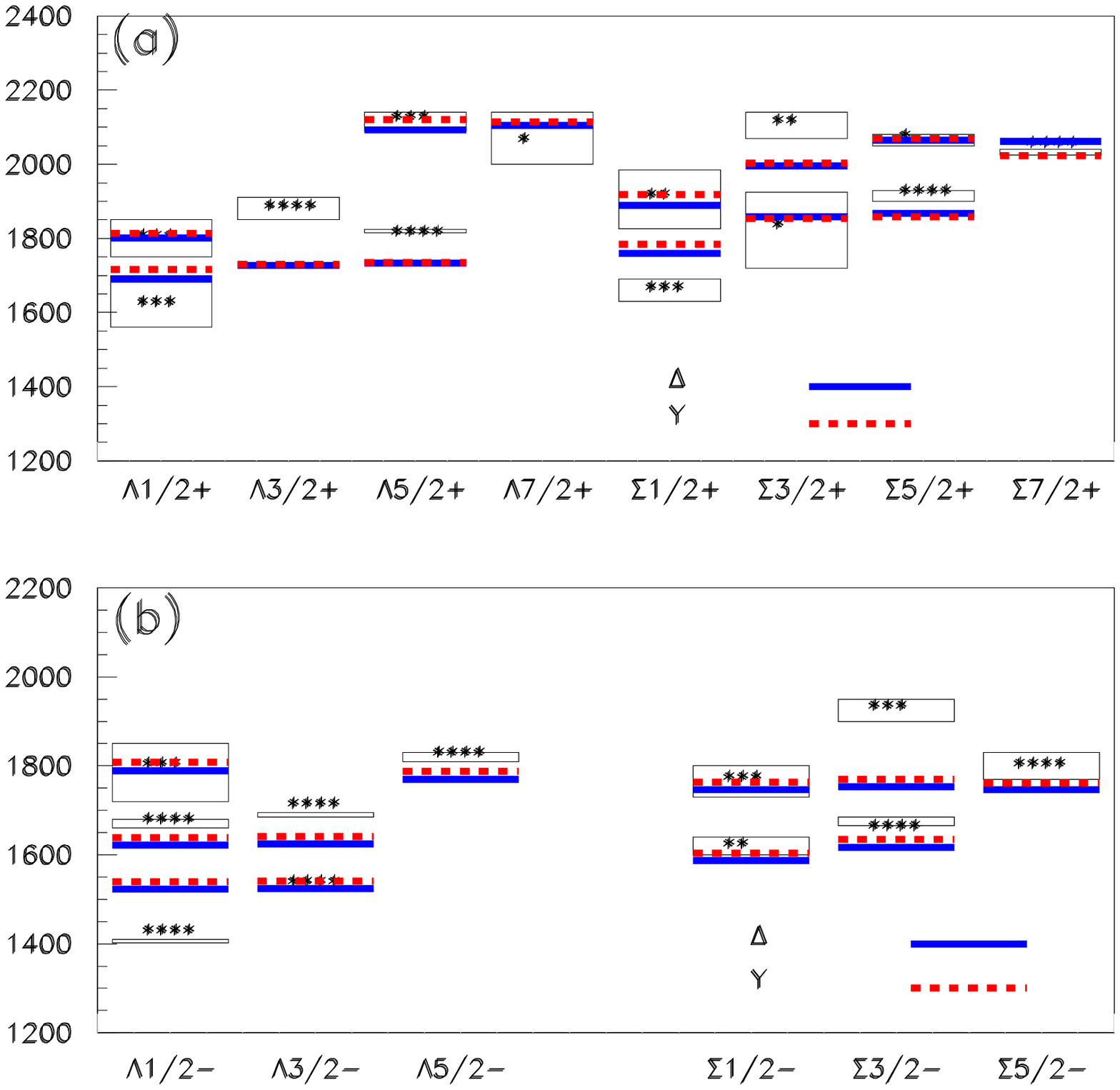}} \hfill
\hglue 0.7cm
\parbox{.47\textwidth}{\epsfysize=6.5cm \epsffile{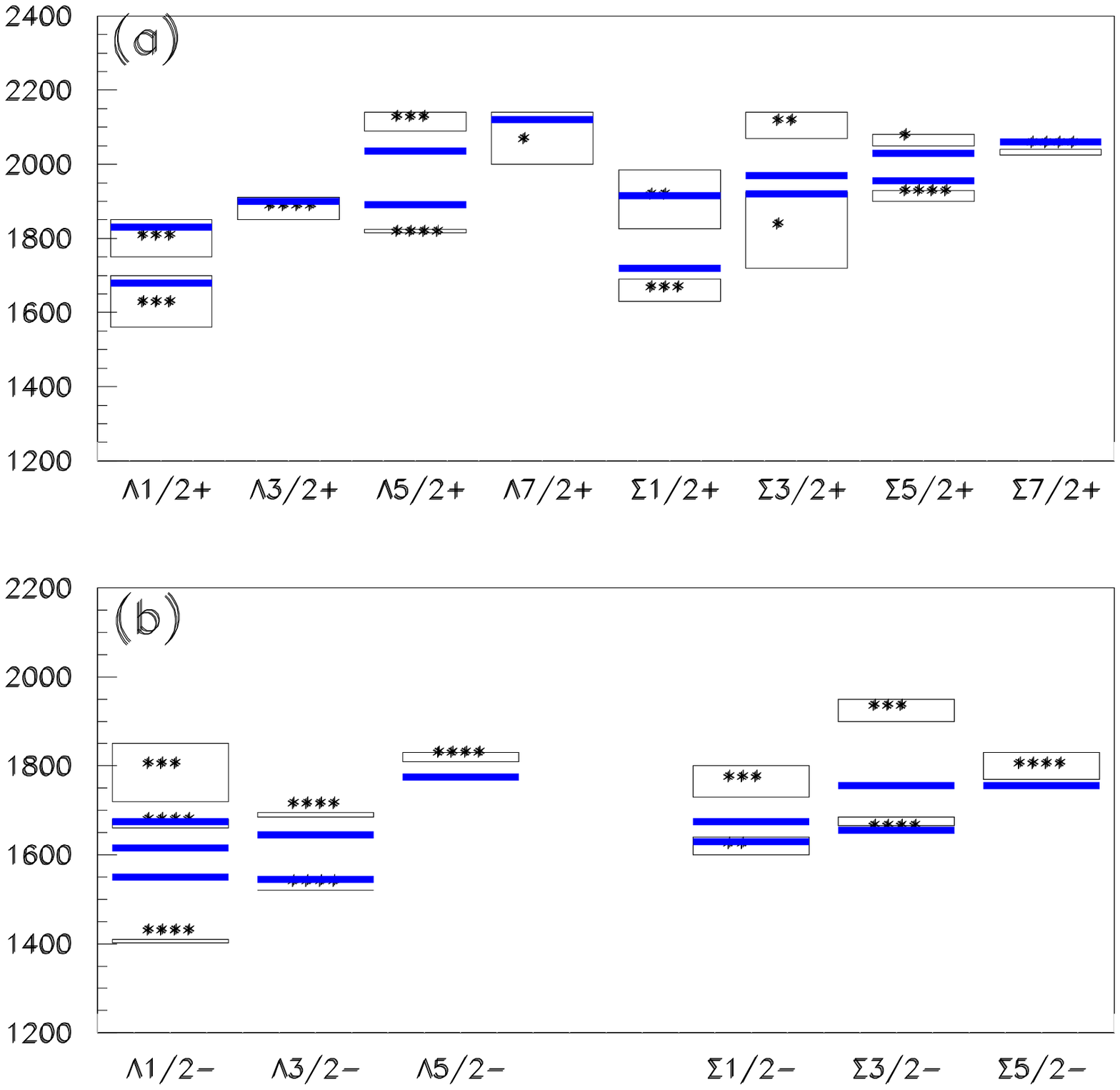}}\hfill
\vspace{0.5cm}
\parbox{.46\textwidth}{\caption{\label{fig:mixrelsposneg}
{\small {The S=-1 baryon spectra in N $\leq$ 2 bands in the chiral
SU(3) quark model. Legend as in
Fig.\ref{fig:mixrenposneg}.}}}}\hfill
\vspace{-3.5cm} 
\parbox{.49\textwidth}{\caption{\label{fig:IsgurLSposneg}
{\small {The S=-1 baryon spectra in N $\leq$ 2 bands in
\cite{Isgur86}. Legend as in Fig.\ref{fig:mixrenposneg}.}}}}\hfill
\vglue 4.5cm
\parbox{.45\textwidth}{\epsfysize=6.5cm\epsffile{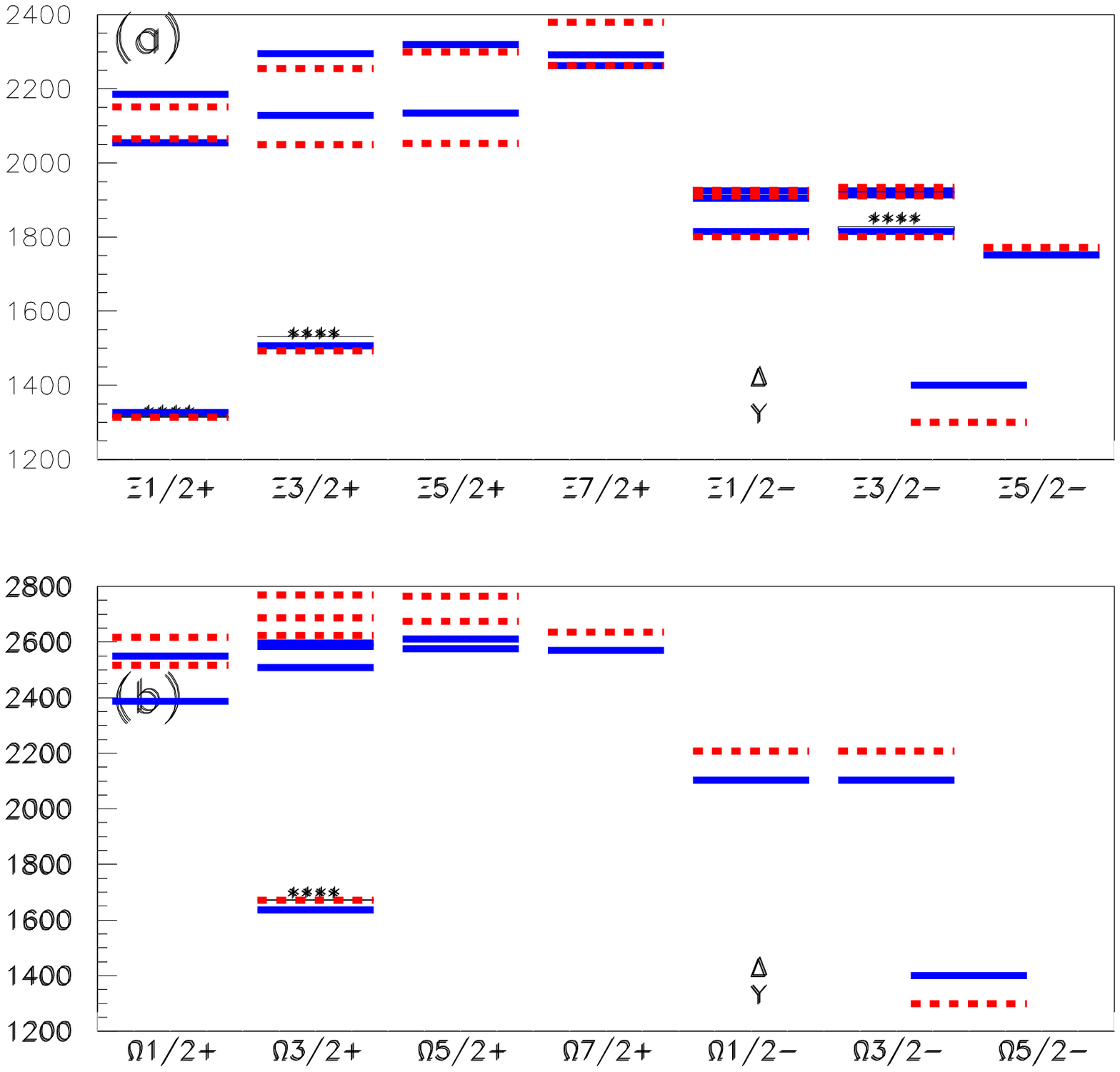}}\hfill
\parbox{.49\textwidth}{\caption{\label{fig:Rekasomeg}
{\small {The S=-2,-3 baryon spectra in N $\leq$ 2 bands in the
chiral
SU(3) quark model. Legend as in Fig.\ref{fig:mixrenposneg}. }}}}
\end{center}
\end{figure}

\tabcolsep 0.1in \renewcommand\arraystretch{0.7}
\begin{table}
\begin{small}
\begin{center}
\caption{\label{tab:mixlspos}{\small {Positive-parity S=-1 baryon
masses in N$\leq$2 bands (in $MeV$). Experimental data are taken
from \cite{PDG}.}}} \vspace{0.3cm}
\begin{tabular}{ccccccc}
\hline\hline
State& $\Delta$-mode&Y-mode&experimental data\\
\hline
$\Lambda^{*}\frac{1}{2}^{+}$& 1115&1115&1115.6&****\\
 & 1690&1717&1560-1700&***\\
 & 1801&1814&1750-1850&***\\
 & 1844&1869& &\\
 & 2088&2079& &\\
 & 2127&2128& &\\
 & 2173&2239& &\\
\hline
$\Lambda^{*}\frac{3}{2}^{+}$ 1792&1727&1730&1850-1910&****\\
 & 2039&2066& &\\
 & 2053&2072& &\\
 & 2098&2090& &\\
 & 2139&2196& &\\
 & 2230&2265& &\\
 & 2414&2323& &\\
\hline
$\Lambda^{*}\frac{5}{2}^{+}$&1733&1735&1815-1825&****\\
 & 2092&2121&2090-2140&***\\
 & 2133&2187& &\\
 & 2162&2207& &\\
 & 2414&2423& &\\
\hline
$\Lambda^{*}\frac{7}{2}^{+}$& 2105&2115&2000-2140&*\\
\hline
$\Sigma^{*}\frac{1}{2}^{+}$& 1192&1192&1192.6&****\\
 & 1759&1784&1630-1690&***\\
 & 1889&1918&1738-1790&*\\
 & 1992&1968&1826-1985&**\\
 & 2064&2060& &\\
 & 2103&2092& &\\
 & 2139&2220& &\\
\hline
$\Sigma^{*}\frac{3}{2}^{+}$& 1371&1370&1382&****\\
 & 1859&1854&1800-1925&*\\
 & 1995&2003&2070-2140&**\\
 & 2056&2060& &\\
 & 2079&2092& &\\
 & 2094&2096& &\\
 & 2103&2124& &\\
 & 2135&2212& &\\
 & 2200&2221& &\\
$\Sigma^{*}\frac{5}{2}^{+}$& 1868&1859&1900-1930&****\\
 & 2065&2069&2050-2080&*\\
 & 2099&2121& &\\
 & 2163&2217& &\\
 & 2199&2237& &\\
\hline
$\Sigma^{*}\frac{7}{2}^{+}$& 2062&2023&2025-2040&****\\
 & 2114&2204& &\\
 \hline\hline
\end{tabular}
\end{center}
\end{small}
\end{table}

\tabcolsep 0.1in \renewcommand\arraystretch{0.7}
\begin{table}
\begin{small}
\begin{center}
 \caption{\label{tab:mixlsneg}{\small {Negative-parity S=-1 baryon
 masses in N=1 bands (in $MeV$).
 Experimental data are taken from \cite{PDG}.}}} \vspace{0.3cm}
\begin{tabular}{ccccccc}
 \hline\hline
State&$\Delta$-mode&Y-mode&experimental data&\\
\hline
$\Lambda^{*}\frac{1}{2}^{-}$& 1523&1540&1402-1410&****\\
 &1622&1639&1660-1680&****\\
 &1789&1808&1720-1850&****\\
\hline
$\Lambda^{*}\frac{3}{2}^{-}$&1524&1541&1518-1520&****\\
 & 1624&1641&1685-1695&****\\
 & 1748&1767& 2307-2372&*\\
\hline
$\Lambda^{*}\frac{5}{2}^{-}$&1769&1788&1810-1830 &****\\
\hline\hline

$\Sigma^{*}\frac{1}{2}^{-}$& 1587&1604&1600-1640&**\\
 &1747&1763&1730-1800&***\\
 &1774&1791&1755-2004&*\\
\hline
$\Sigma^{*}\frac{3}{2}^{-}$&1617&1635&1578-1584&**\\
 & 1743&1760&1665-1685&****\\
 & 1753&1769& 1900-1950&***\\
\hline
$\Sigma^{*}\frac{5}{2}^{-}$& 1746&1762&1770-1780 &****\\
\hline\hline
\end{tabular}
\end{center}
\end{small}
\end{table}

The baryon spectra with strange number S=~-1 in N $\leq$ 2 bands are
given in Table~\ref{tab:mixlspos}-\ref{tab:mixlsneg} and
Fig.\ref{fig:mixrelsposneg}. Once again, the results of Ref.
\cite{Isgur86} are pictured in Fig. \ref{fig:IsgurLSposneg}. From
Fig. \ref{fig:mixrelsposneg}, one sees that the fitting quality of
spectra to the experimental data in this sector is not as good as
that in the non-strange sector, but most of the resultant resonances
are located within experimental error bars. In general, the fitting
quality is similar to that in Ref. \cite{Isgur86}. The spectra with
the $\Delta$-mode confining potential is close to those with the
Y-mode, just as that in the non-strange sector. The evaluated mass
of $\Lambda^{*}({\frac{1}{2}^{-}},1405)$ is more than 100 MeV larger
than the experimental value when the spin-orbit interaction, which
causes the mass splitting between
$\Lambda^{*}({\frac{1}{2}^{-}},1405)$ and
$\Lambda^{*}({\frac{3}{2}^{-}},1520)$, is omitted in the
calculation. Inclusion of such an interaction would result in a more
reasonable spectrum for strange baryons Ref. \cite{YuYW}.

\tabcolsep 0.1in \renewcommand\arraystretch{0.7}
\begin{table}
\begin{small}
\begin{center}
\caption{\label{tab:ogeksom}{\small {$\Xi$ and $\Omega$ baryon
masses in N$\leq$2 bands in the chiral SU(3) quark model (in $MeV$).
Experimental data are taken from \cite{PDG}.}}} \vspace{0.3cm}
\begin{tabular}{ccccccccccc}
\hline\hline
\multicolumn{1}{c}{State}&\multicolumn{1}{c}{Model}&\multicolumn{8}{c}{
Predicted masses (MeV)}\\
\hline
$\Xi^{*}\frac{1}{2}^{+}$&$\triangle$-shape&1327&1894&2016&2122&2190&2234&2263& &\\
 &$Y$-shape&1314&1907&2028&2085&2175&2212&2325\\
$\Xi^{*}\frac{3}{2}^{+}$&$\triangle$-shape&1506&1947&2122&2185&2204&2215&2234&2246&2361\\
 &$Y$-shape&1493&1927&2115&2176&2204&2212&2232&2312&2367\\
$\Xi^{*}\frac{5}{2}^{+}$&$\triangle$-shape&1952&2191&2218&2268&2361\\
 &$Y$-shape&1932&2177&2230&2327&2366\\
$\Xi^{*}\frac{7}{2}^{+}$&$\triangle$-shape&2184&2233& &\\
 &$Y$-shape&2134&2304\\
 \hline
$\Omega^{*}\frac{1}{2}^{+}$&$\triangle$-shape&2262&2326\\
 &$Y$-shape&2516&2618\\
$\Omega^{*}\frac{3}{2}^{+}$&$\triangle$-shape&1636&2241&2339&2347\\
 &$Y$-shape&1672&2623&2688&2770\\
$\Omega^{*}\frac{5}{2}^{+}$&$\triangle$-shape&2333&2365\\
 &$Y$-shape&2674&2766\\
$\Omega^{*}\frac{7}{2}^{+}$ &$\triangle$-shape&2322&\\
 &$Y$-shape&2636\\
\hline

$\Xi^{*}\frac{1}{2}^{-}$&$\triangle$-shape&1756&1871&1890\\
 &$Y$-shape&1760&1875&1896\\
$\Xi^{*}\frac{3}{2}^{-}$& $\triangle$-shape&1756&1871&1896\\
 &$Y$-shape&1760&1875&1901\\
$\Xi^{*}\frac{5}{2}^{-}$&$\triangle$-shape&1892\\
 &$Y$-shape&1898\\
\hline

$\Omega^{*}\frac{1}{2}^{-}$&$\triangle$-shape&2104\\
 &$Y$-shape&2099\\
$\Omega^{*}\frac{3}{2}^{-}$&$\triangle$-shape&2104\\
 &$Y$-shape&2099\\
\hline\hline
\end{tabular}
\end{center}
\end{small}
\end{table}
The S= -2 and -3 baryon spectra in $N\leq 2$ bands are given in Fig.
\ref{fig:Rekasomeg} and Table ~\ref{tab:ogeksom}.  The deviations of
the results with the Y-mode from those with $\Delta$-mode are
similar to those in the former sectors. The results show that both
the $\Delta$-mode and the Y-mode confining potentials can lead to
reasonable baryon spectra.
\subsection{Comparison of $\Delta$-mode and Y-mode }
\begin{figure*}
\begin{center}
\parbox{.47\textwidth}{\epsfysize=6.5cm\epsffile{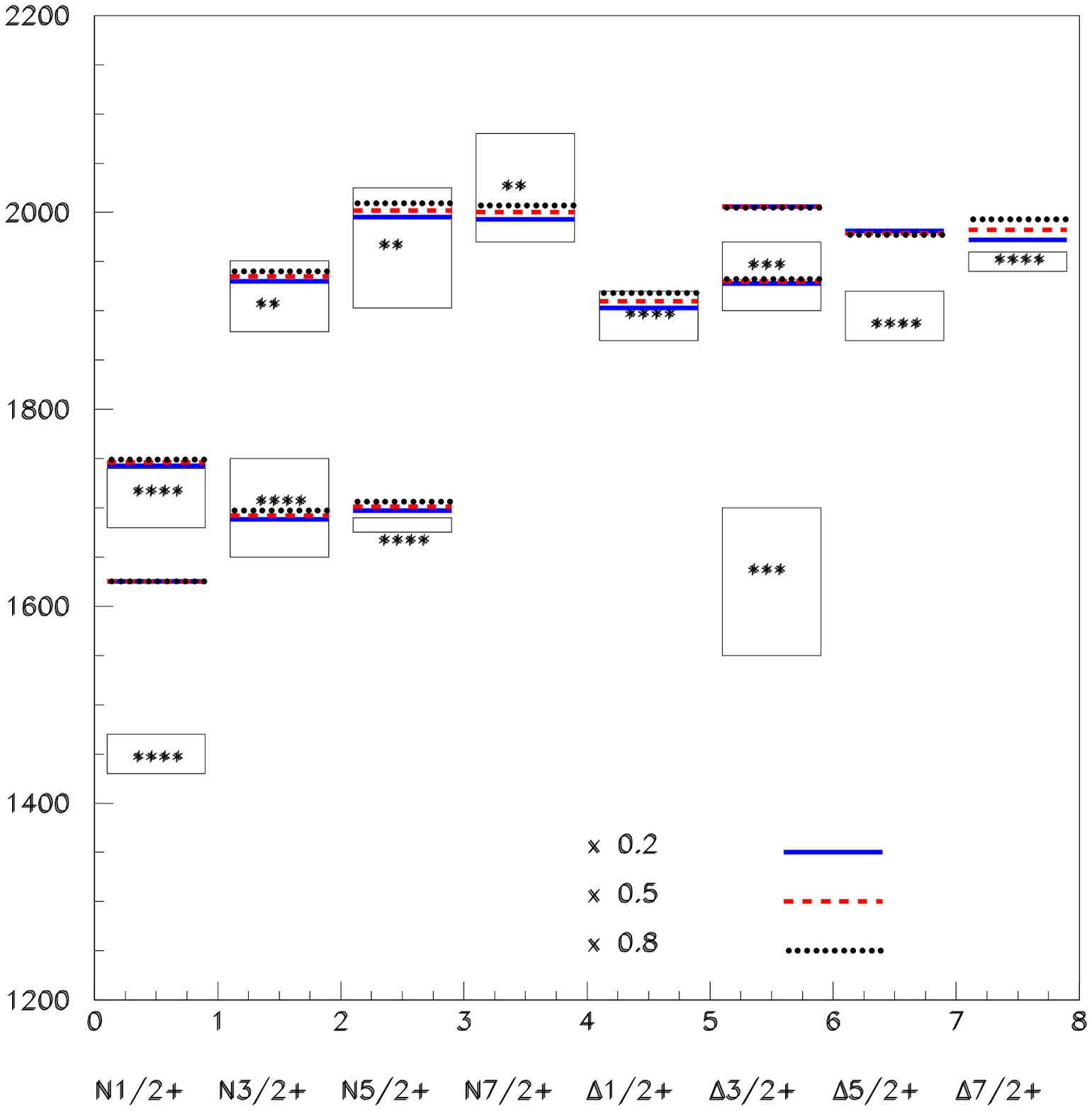}} \hfill
\hglue 0.7cm
\parbox{.47\textwidth}{\epsfysize=6.5cm \epsffile{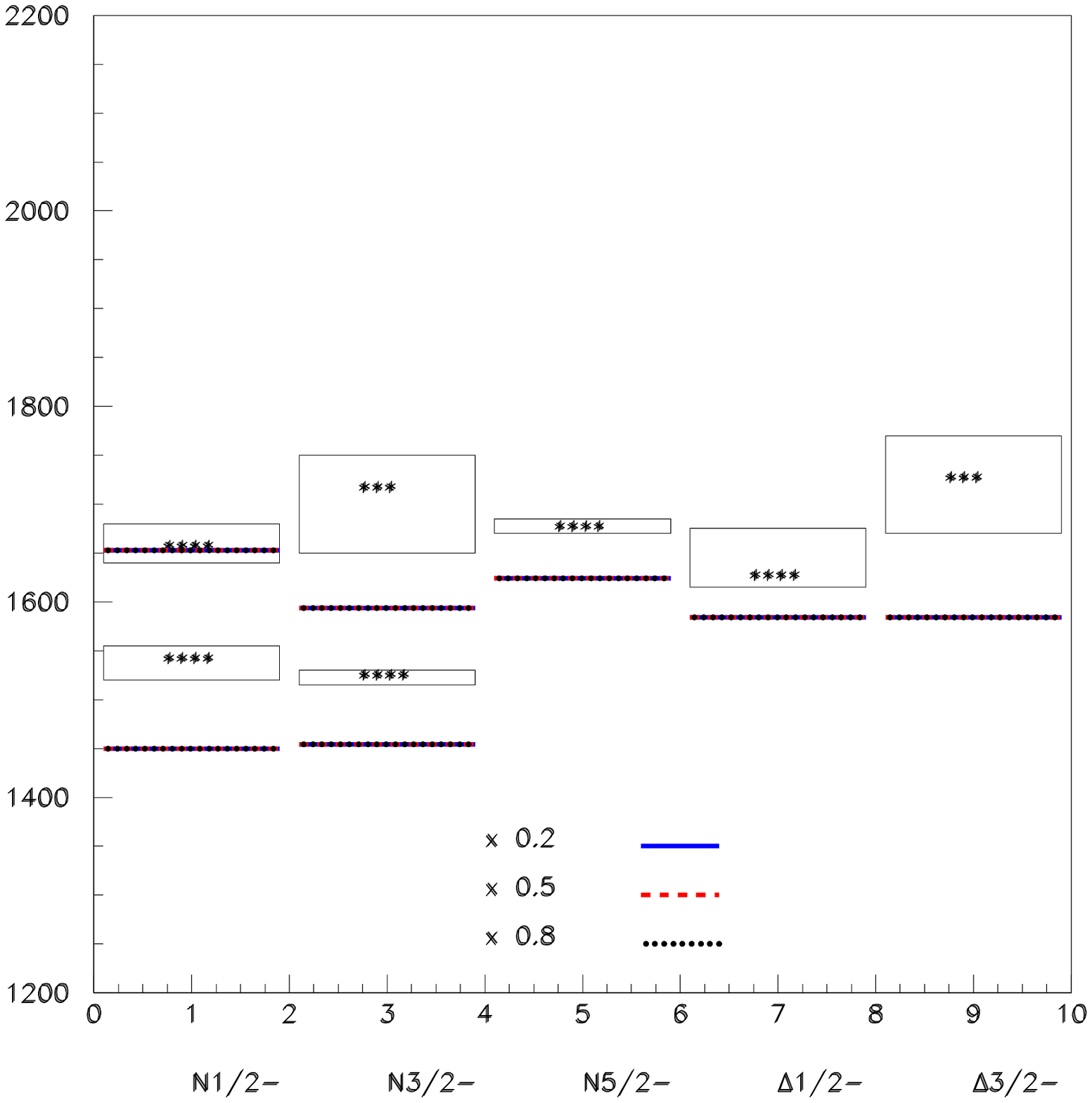}}\hfill
\parbox{.95\textwidth}{\caption{\label{fig:2b3bN}
{\small {S=0 baryon spectra with mixed $\Delta$- and Y-mode
confining potentials in $N\leq 2$ bands. The solid, dashed and
dotted bars correspond to the results with x being 20\%, 50\% and
80\%, respectively.}}}}
\end{center}
\end{figure*}

\begin{figure*}
\begin{center}
\parbox{.47\textwidth}{\epsfysize=6.5cm\epsffile{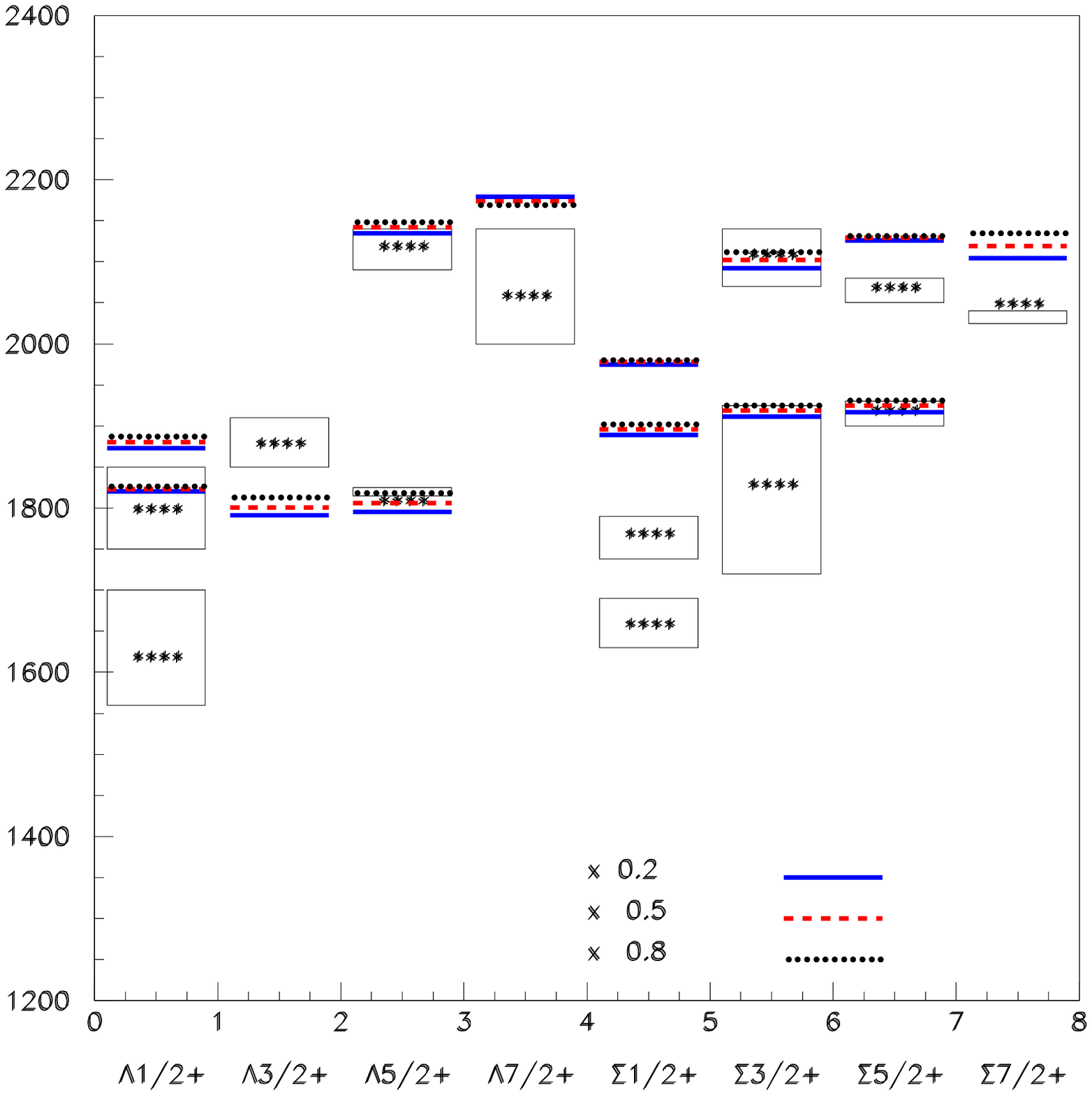}} \hfill
\hglue 0.7cm
\parbox{.47\textwidth}{\epsfysize=6.5cm \epsffile{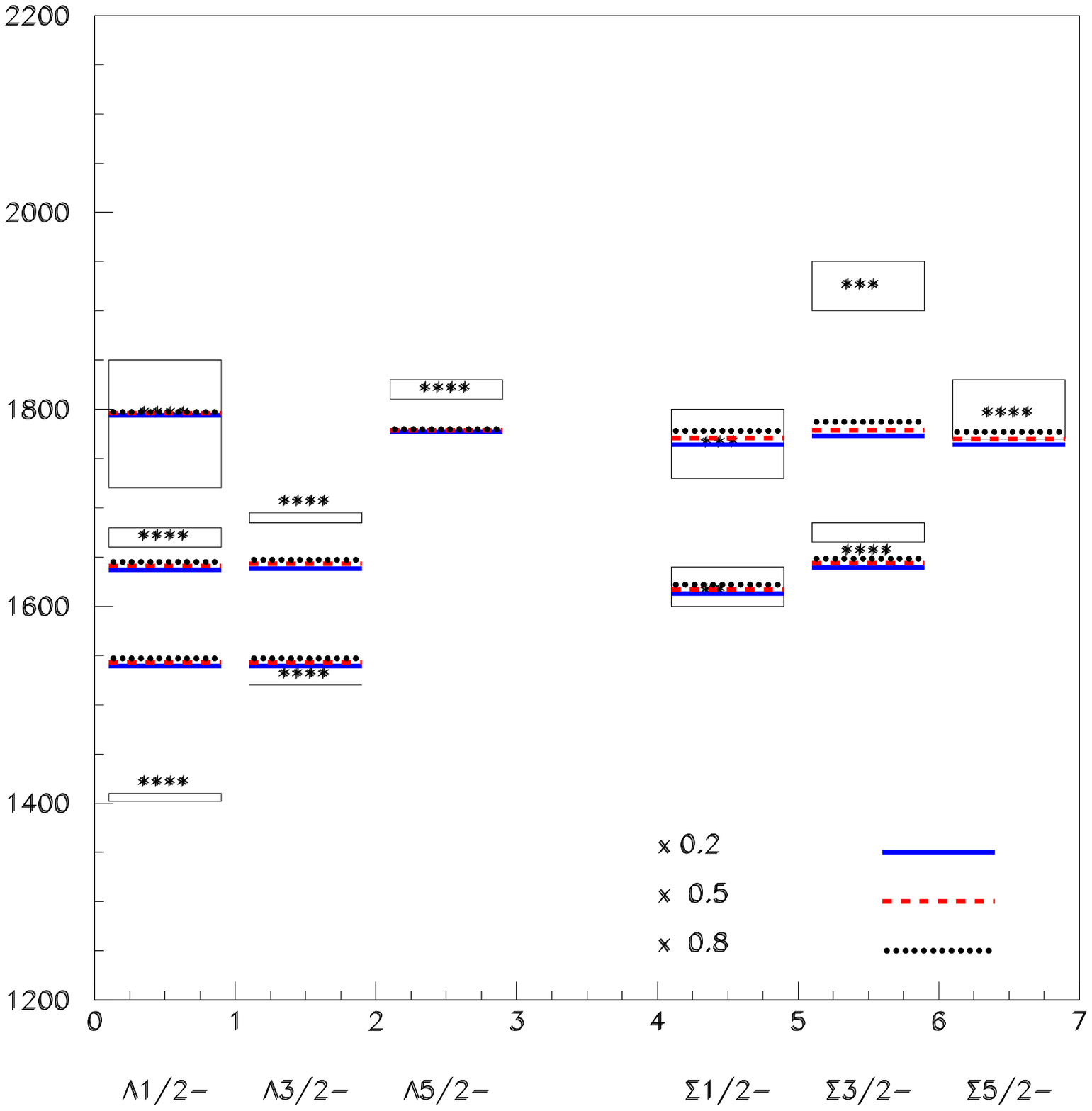}}\hfill
\parbox{.95\textwidth}{\caption{\label{fig:2b3bS}
{\small {S=-1 baryon spectra with mixed $\Delta$- and Y-mode
confining potentials in $N\leq 2$ bands. The solid, dashed and
dotted bars correspond to the results with x being 20\%, 50\% and
80\%, respectively. }}}}
\end{center}
\end{figure*}

In order to study the effect of different confinement modes on the
baryon spectrum, we now use a confining potential in which both
$\Delta$-mode and the Y-mode are included
\begin{eqnarray}
V^{conf}=xV^{conf}_{\triangle}+(1-x)V^{conf}_{Y},
\end{eqnarray}
where x stands for the fraction of the $\Delta$-mode in the whole
confining potential. The S=0 and -1 baryon spectra with x values of
0.2, 0.5 and 0.8 are plotted in Figs. \ref{fig:2b3bN} and
\ref{fig:2b3bS}, respectively. The resultant baryon spectra with
different x-values show very small differences.

\begin{figure*}
\begin{center}
\begin{center}
 \epsfysize=4.5cm\hspace*{0.2cm}\vspace*{0.2cm}
\epsffile{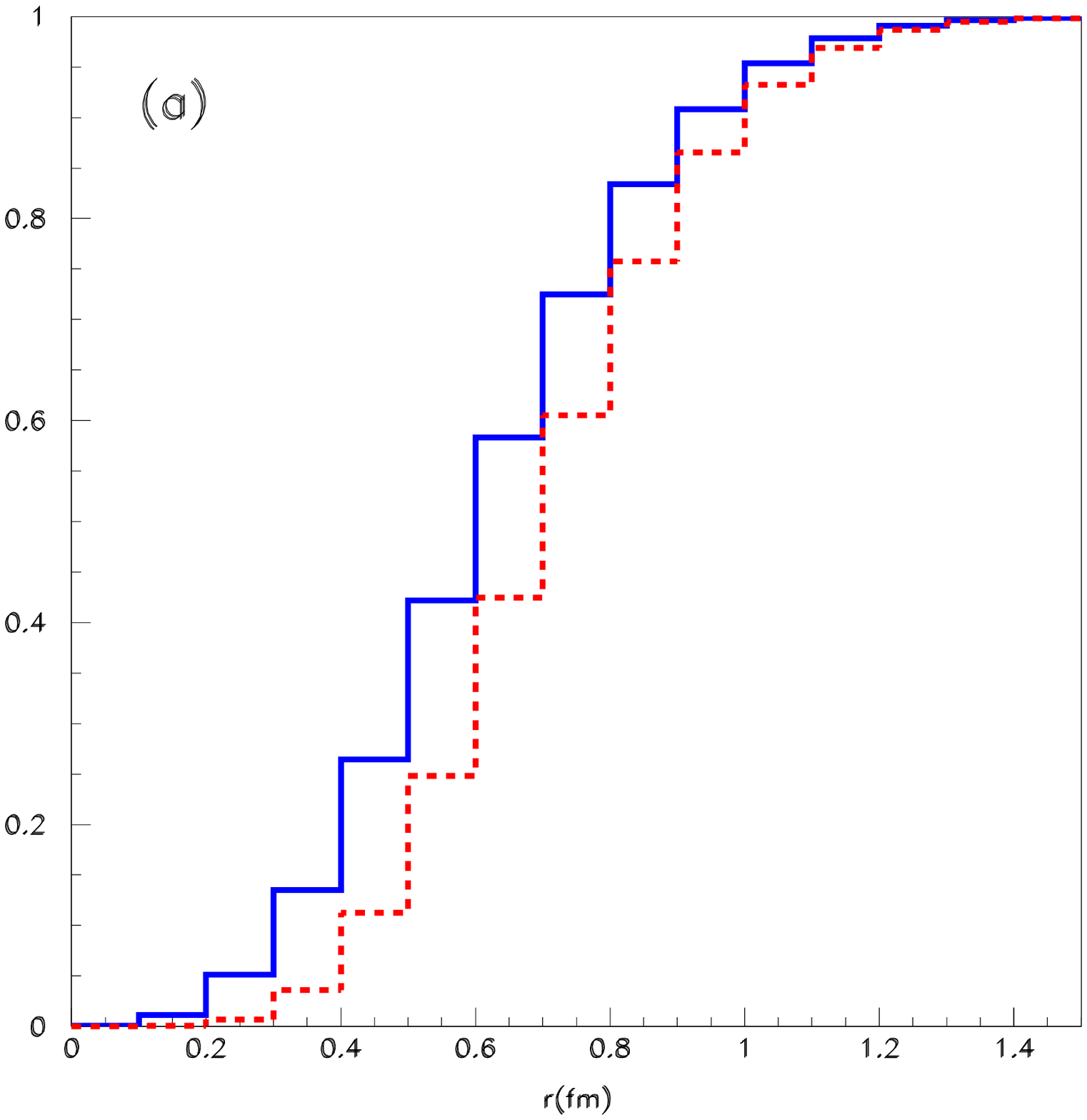}
 \epsfysize=4.5cm\hspace*{0.2cm}\vspace*{0.2cm}
\epsffile{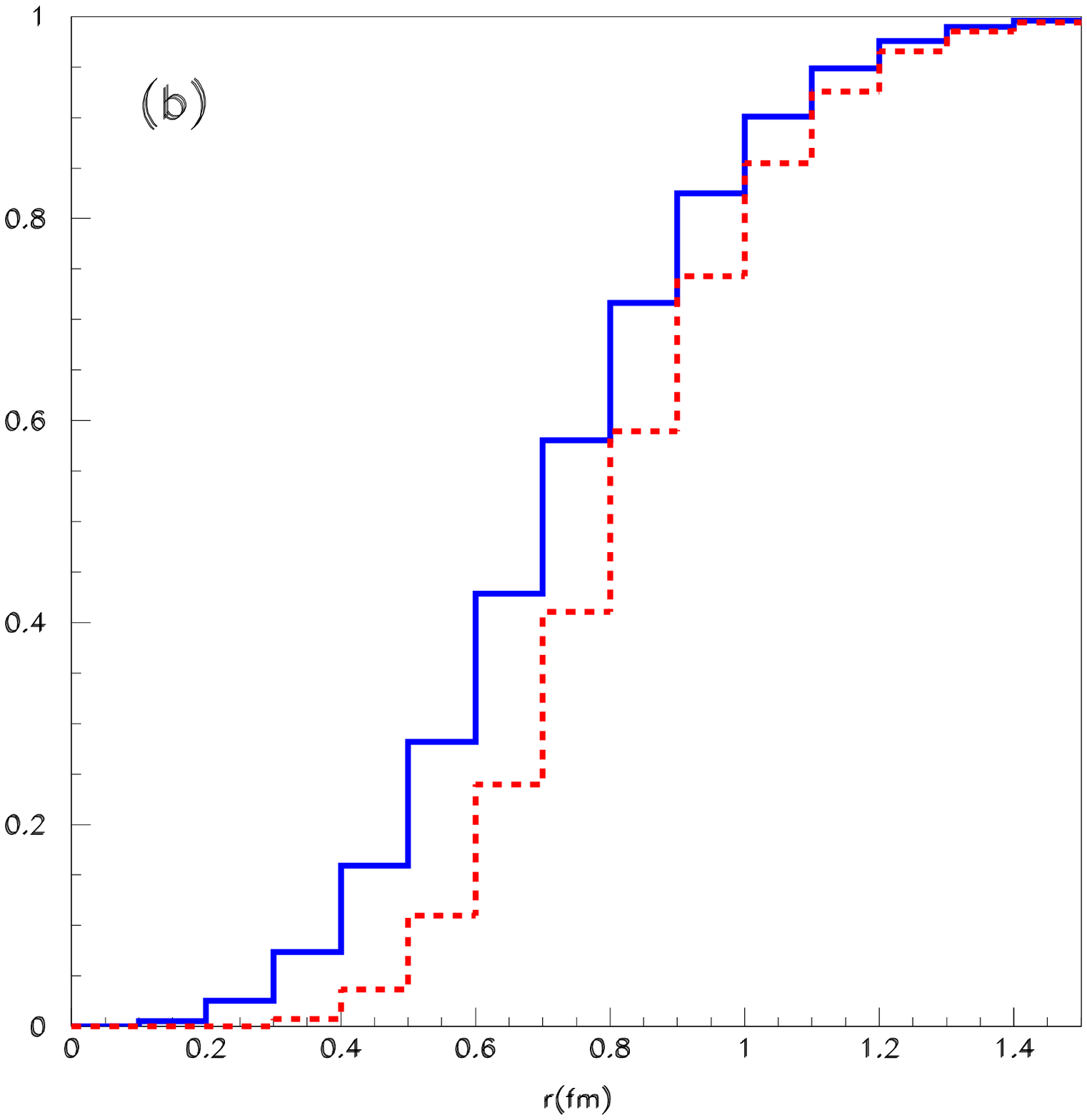}
 \epsfysize=4.5cm\hspace*{0.2cm}\vspace*{0.2cm}
 \epsffile{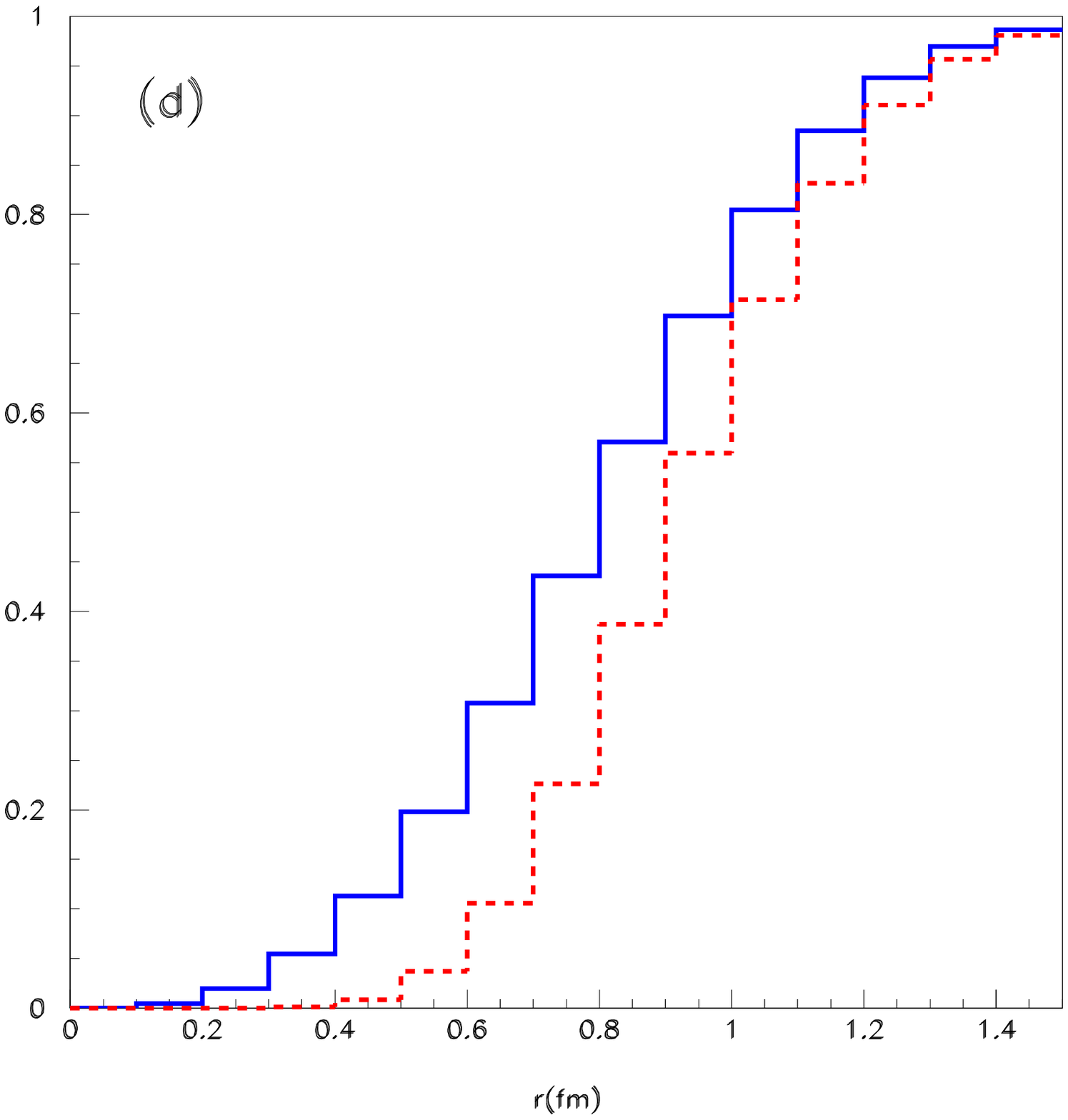}
\parbox{\textwidth}{\caption{\label{fig:compare}
{\small {Contributions of the matrix elements of $\Delta$-mode and
Y-mode confining potentials with respect to the quark separation.
(a) $|^{2}N(56,0^{+})\frac{1}{2}^{+}>$ state, (b)
$|^{2}N(70,1^{-})\frac{1}{2}^{-}>$ state and (c)
$|^{2}N(56,2^{+})\frac{1}{2}^{+}>$ state with the notation
$|^{2J+1}N(N_6,N^{P})J^{P}>$, where $J$, $N_6$, $N$ and $P$ denote
the total angular momentum, the dimension of $SU_{SF}(6)$ group, the
principal quantum number and the parity of the state, respectively.
The solid and dashed curves correspond to the $\Delta$-mode
confining potential and the Y-mode confining potential cases,
respectively. }}}}
\end{center}
\end{center}
\end{figure*}

From the structure of the flux-tube model, the difference between
the Y-mode and the $\Delta$-mode potentials comes from different
geometric shapes of flux-tubes in baryon. By denoting $L_{Y}$ and
$L_{\Delta}$ as the lengths of the Y-mode and $\Delta$-mode
flux-tubes, respectively, we have
\begin{eqnarray}
\frac{L_{\Delta}}{2}\leq L_{Y}\leq
\frac{2}{\sqrt{3}}\frac{L_{\Delta}}{2},
\end{eqnarray}
where the left equal sign means that the junction point is sitting
at the apex and the right equal sign represents the fact that three
quarks in baryon stay at apexes of an equilateral triangle. Further
considering a factor of $\frac{1}{2}$ in the string tension of the
$\Delta$-mode confining potential, the ratio of the Y-mode and
$\Delta$-mode confining potentials satisfies
\begin{eqnarray}
1\leq\frac{L_{Y}}{L_{\Delta}/2}(=\frac{V_Y^{conf}}{V_{\Delta}^{conf}})\leq
\frac{2}{\sqrt{3}}\simeq 1.15~.
\end{eqnarray}
This relation indicates that the maximal difference between the
Y-mode and $\Delta$-mode confining potentials is 15\%. Because the
calculated baryon mass is closely related to the averaged values of
potentials, one might not be able to distinguish the effects from
two confining modes by studying the spectrum only. However, by
carefully investigating the matrix element (ME) of the confining
potential in different integrating intervals, one might see the
different effects from different confining modes. In
Fig.~\ref{fig:compare}, we demonstrate the values of the matrix
element of confining potentials in the Y-mode and $\Delta$-mode with
respect to the quark separation step by step in several states. In
this figure, the matrix element at $r=\rho_{max}=\lambda_{max}$ with
$\rho_{max}$ and $\lambda_{max}$ being the upper integrating limit
of the $\rho$ and $\lambda$ integrations are normalized to 1 for an
easy comparison. Evidently, at the short and medium distances
$(0.2fm-0.8fm)$, the $\Delta$-mode is dominant and at the large
distances the Y-mode provides more contributions. This result
coincides with Alexandrou's argument \cite{Alexandrou}. Moreover, in
lower lying states, the increase rate of the matrix element of the
Y-mode potential is much slower than that of the $\Delta$-mode
potential in the distances less than $0.6fm$, but becomes much
faster as $r>0.8fm$, especially in higher level states.

\section{The extended chiral SU(3) quark model}

Recently Zhang {\it et al} extended the chiral SU(3) quark model by
including the exchanges of the singlet and octet vector meson fields
between quarks. In this extended model, the vector meson induced
quark-quark interaction is introduced through the interaction
Lagrangian
\begin{eqnarray}
{\cal L}_I^v = -g_{chv} \bar{\psi}\gamma_\mu T^a A^\mu_a \psi
-\frac{f_{chv}}{2M_P} \bar{\psi} \sigma_{\mu\nu} T^a \partial^\nu
A^\mu_a \psi,
\end{eqnarray}
where $g_{chv}$ and $f_{chv}$ are the coupling constants for the
vector coupling and the tensor coupling, respectively. The form of
the additional potential can be written as
\begin{eqnarray}
V^{v}_{ij} = \sum_{a=0}^8 V_{\rho_a}({\bm r}_{ij}),
\end{eqnarray}
where $\rho_{0},..,\rho_{8}$ denote the singlet and octet vector
fields, and
\begin{eqnarray}
V_{\rho_a}({\bm r}_{ij})&=&C(g_{chv},m_{\rho_a},\Lambda)\left\{
X_1(m_{\rho_a},\Lambda,r_{ij})+ \frac{m^2_{\rho_a}}{6m_{q_i}m_{q_j}}
\left(1+\frac{f_{chv}}{g_{chv}}\frac{m_{q_i}+m_{q_j}}{M_P}+\frac{f^2_{chv}}{g^2_{chv}}
\right. \right. \nonumber \\
&& \left. \left. \times \frac{m_{q_i}m_{q_j}}{M^2_P}\right)
X_2(m_{\rho_a},\Lambda,r_{ij})({\bm \sigma}_i\cdot{\bm \sigma}_j)
\right\}[\lambda_a(i)\lambda_a(j)] + V_{\rho_a}^{\bm {l \cdot
s}}({\bm r}_{ij}) + V_{\rho_a}^{ten}({\bm r}_{ij}),
\end{eqnarray}
\begin{eqnarray}
V_{\rho_a}^{\bm {l \cdot s}}({\bm r}_{ij})&=&
-C(g_{chv},m_{\rho_a},\Lambda)\frac{3m^2_{\rho_a}}{4m_{q_i}m_{q_j}}
\left(1+\frac{f_{chv}}{g_{chv}}\frac{2(m_{q_i}+m_{q_j})}{3M_P}\right)
\nonumber \\
&&\times
\left\{G(m_{\rho_a}r_{ij})-\left(\frac{\Lambda}{m_{\rho_a}}\right)^3
G(\Lambda r_{ij})\right\}[{\bm L \cdot ({\bm \sigma}_i+{\bm
\sigma}_j)}][\lambda_a(i)\lambda_a(j)],
\end{eqnarray}
\begin{eqnarray}
V_{\rho_a}^{ten}({\bm r}_{ij}) &=& -C(g_{chv},m_{\rho_a},\Lambda)
\frac{m^2_{\rho_a}}{12m_{q_i}m_{q_j}}
\left(1+\frac{f_{chv}}{g_{chv}}\frac{m_{q_i}+m_{q_j}}{M_P}+\frac{f^2_{chv}}{g^2_{chv}}\frac{m_{q_i}m_{q_j}}{M^2_P}\right)
\nonumber \\
&&\times\left\{H(m_{\pi_a}r_{ij})-\left(\frac{\Lambda}{m_{\pi_a}}\right)^3
H(\Lambda r_{ij})\right\}\hat{S}_{ij}[\lambda_a(i)\lambda_a(j)],
\end{eqnarray}
with $m_{\rho_a}$ being the mass of the vector meson.

In the extended model, some additional parameters are introduced
\cite{DaiLR03}. Based on $g_{NN\rho}$ value in the phenomenological
$NN$ interaction model, the coupling constants $g_{chv}$ and
$f_{chv}$ can be estimated as $g_{chv}=2.351$ and
$f_{chv}=2/3g_{chv}$, respectively. The mixing angle between
$\omega_1$ and $\omega_8$ is taken to be $\theta^V=35.26^\circ$,
which indicates an ideal mixtrue. The model parameters are
summarized in Table \ref{tab:Epara}.

{\small
\begin{table}[htb]
\caption{\label{tab:Epara} Parameters of extended chiral SU(3) quark
model. The vector meson masses $m_{\rho}=770MeV$,
 $m_{K^{*}}=892MeV$, $m_{\omega}=782MeV$, $m_{K^{*}}=1020MeV$ and
the cutoff momentum $\Lambda=1100MeV$.
 (1)$f_{chv}=0$, (2)$f_{chv}\neq 0$.}
\begin{center}
\begin{tabular}{cccc}
 \hline\hline
 \multicolumn{1}{c}{}&\multicolumn{2}{c}{Extended chiral SU(3) quark model}\\
 & $~~~~~~~~~~~~~~~~~~~~~~$(1) & (2) \\
 \hline
$\omega(MeV)$&$~~~~~~~~~~~~~~~~~~~~~~$497.6&497.6\\
$g_{ch}$&$~~~~~~~~~~~~~~~~~~~~~~$2.621&2.621\\
$g_{chv}$&$~~~~~~~~~~~~~~~~~~~~~~$2.351&1.972\\
$f_{chv}/g_{chv}$&$~~~~~~~~~~~~~~~~~~~~~~$--&2/3\\
$m_{\sigma}(MeV)$&$~~~~~~~~~~~~~~~~~~~~~~$535&547\\
$g_{u}$&$~~~~~~~~~~~~~~~~~~~~~~$0.521&0.598\\
$g_{s}$&$~~~~~~~~~~~~~~~~~~~~~~$0.545&0.614\\
$b_{uu}(GeV/fm)$&$~~~~~~~~~~~~~~~~~~~~~~$ 0.57&0.58\\
$b_{us}(GeV/fm)$&$~~~~~~~~~~~~~~~~~~~~~~$ 1.03&0.99\\
$b_{ss}(GeV/fm)$&$~~~~~~~~~~~~~~~~~~~~~~$ 2.01&2.04\\
$c_{uu}(MeV)$&$~~~~~~~~~~~~~~~~~~~~~~$ -566&-524\\
$c_{us}(MeV)$&$~~~~~~~~~~~~~~~~~~~~~~$ -855&-769\\
$c_{ss}(MeV)$&$~~~~~~~~~~~~~~~~~~~~~~$ -1285&-1224\\
 \hline\hline
\end{tabular}
\end{center}
\end{table}}

\begin{figure}
\begin{center}\vspace*{0.5cm}
\parbox{.47\textwidth}{\epsfysize=6.5cm\epsffile{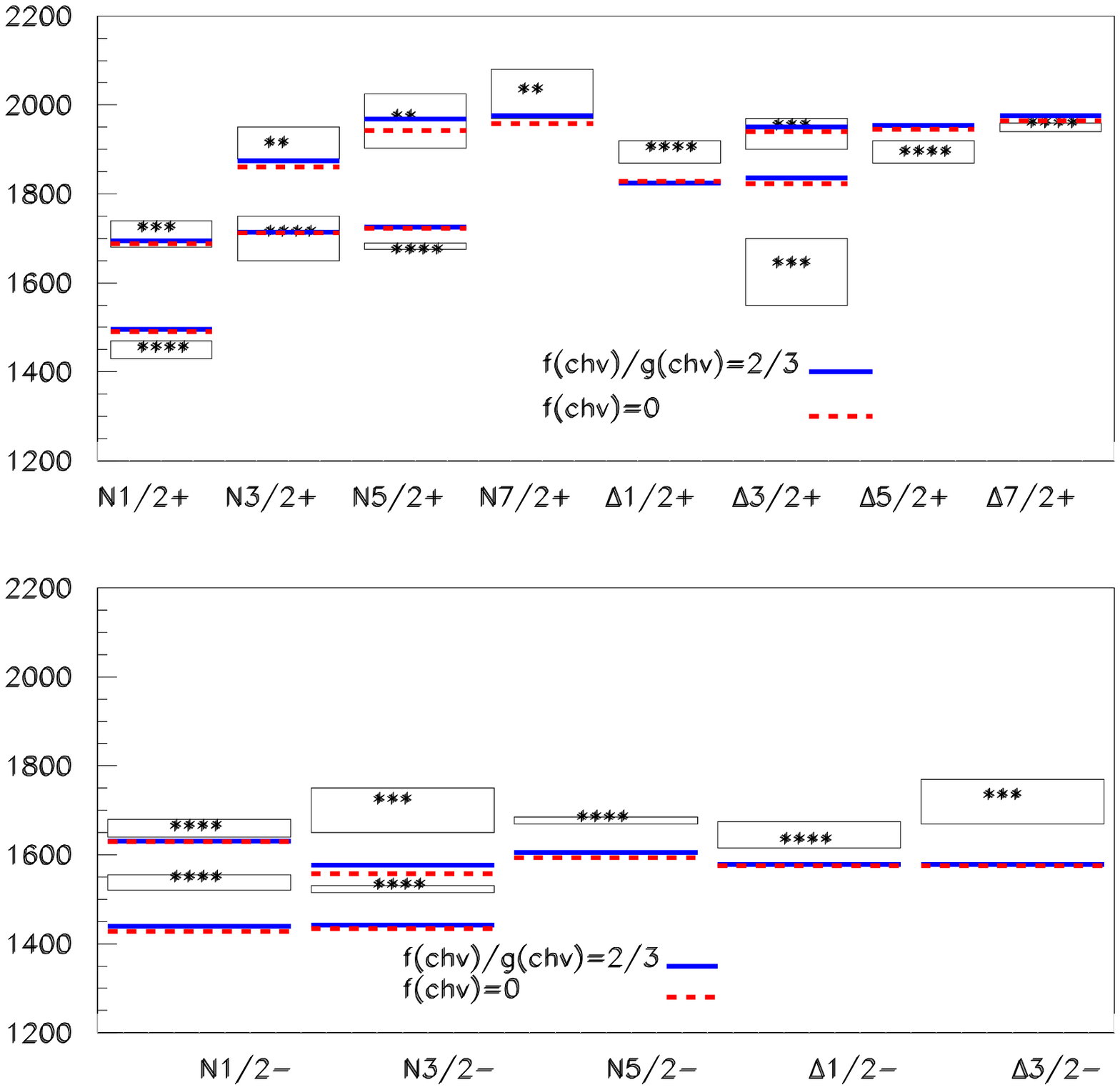}} \hfill
\hglue 0.7cm
\parbox{.47\textwidth}{\epsfysize=6.5cm \epsffile{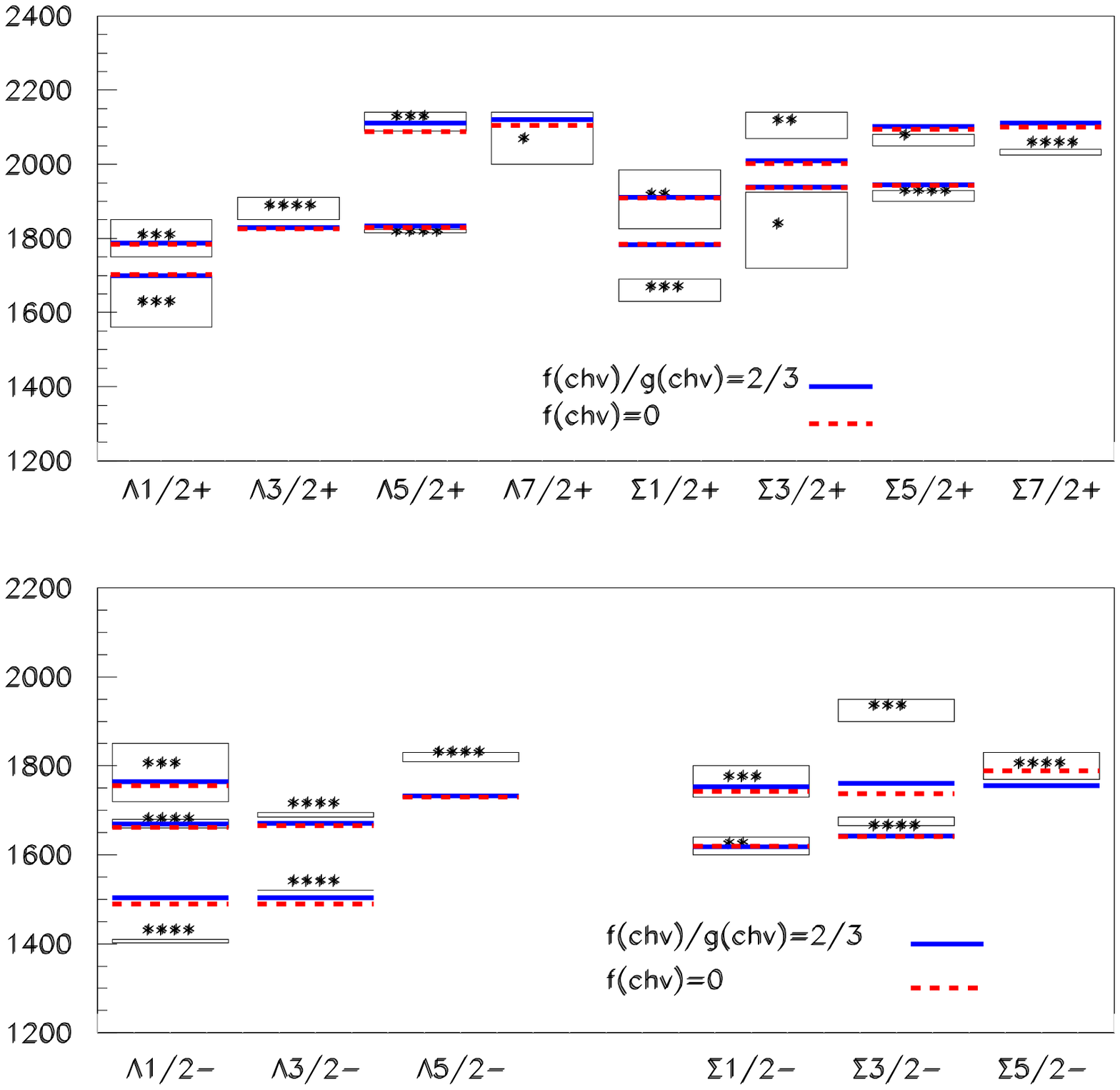}}\hfill
\vspace{0.5cm}
\parbox{.46\textwidth}{\caption{\label{fig:enposnegvect}
{\small {The non-strange baryon spectra in N $\leq$ 2 bands in the
extended chiral SU(3) quark model. The solid and dashed bars
represent the results with and without the tensor coupling,
respectively.}}}}\hfill
\vspace{-3.5cm} 
\parbox{.49\textwidth}{\caption{\label{fig:elsposnegvect}
{\small {The S=-1 baryon spectra in N $\leq$ 2 bands in the extended
chiral SU(3) quark model. Legend as in
Fig.\ref{fig:enposnegvect}.~~~~~~~~~~~~~~~~~~~~~~~~~~~~~~~~}}}}\hfill
\vglue 4.5cm
\parbox{.45\textwidth}{\epsfysize=6.5cm\epsffile{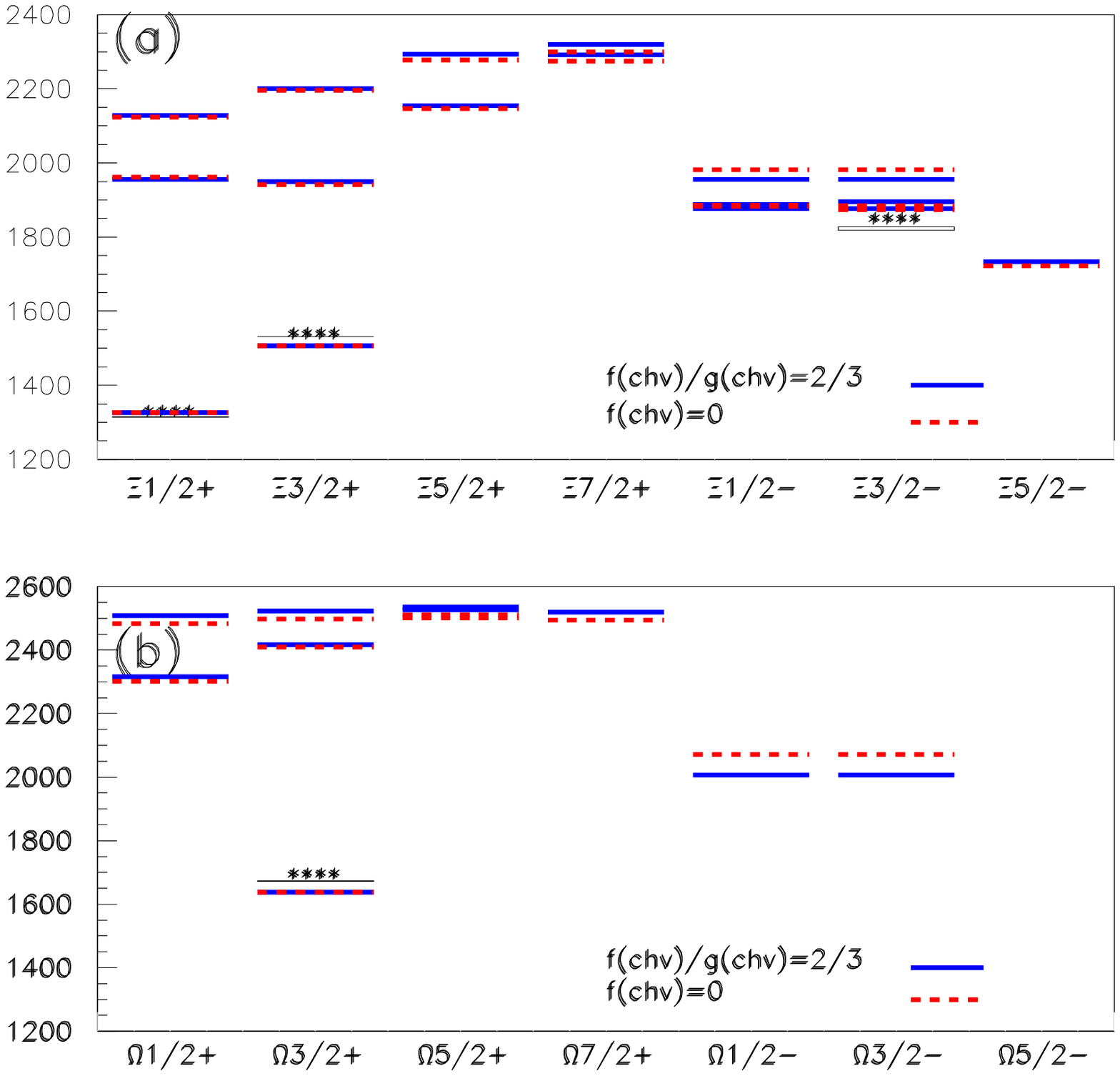}}\hfill
\parbox{.49\textwidth}{\caption{\label{fig:kascaomegadvect}
{\small {The S=-2,-3 baryon spectra in N $\leq$ 2 bands in the
extended chiral SU(3) quark model. Legend as in Fig.\ref{fig:enposnegvect}. }}}}
\end{center}
\end{figure}

In the chiral SU(3) quark model, we find that the baryon spectra in
the $\Delta$-mode and the Y-mode confining potential cases are
rather close, and that one cannot definitely distinguish which
confining mode is better through the spectra only because the mass
of baryon is closely related to the averaged value of the confining
potential operator. Therefore, in order to see whether a reasonable
description for baryon spectra can be obtained in the extended
chiral SU(3) quark model also, we calculate the baryon spectra with
the $\Delta$-mode only. The baryon spectra for S = 0, -1, -2, -3 in
the extended chiral SU(3) quark model are plotted in
Figs.\ref{fig:enposnegvect}-\ref{fig:kascaomegadvect}. In these
figures, the solid and dashed bars denote the results with and
without the tensor coupling in the vector meson potential,
respectively. The fitting quality in the extended model is close to
that in the chiral SU(3) quark model except for a few states, and
the deviations between the states with vector coupling of the vector
meson potential only and those with both vector and tensor couplings
of the vector meson potential are very small. The resultant mass of
Roper resonance ($N(\frac{1}{2}^+,1440)$) is 25 MeV smaller than
that in the chiral SU(3) quark model. It is even closer to the
experimental value, but is still larger than the first orbital
resonance ($N(\frac{1}{2}^-,1535)$). Moreover, in the extended
model, the level intervals among $\Lambda(\frac{1}{2}^-)$ states
become larger.

It should be mentioned that because the effect of the vector meson
exchange can partially replace the role of the one gluon exchange,
introducing vector meson exchange potential can reduce the strong
coupling constant $\alpha_s$ to some extend, which is desirable by
QCD expectation.

\section{nonstrange baryon strong decay}

A successful hadron model should be able to explain as much data as
possible such as the spectrum, magnetic moments, decay width and
etc.. The chiral SU(3) quark model is quite successful in
reproducing the experimental data of the hadron-hadron scattering
and baryon spectrum. Now, we would check if this model can give a
reasonable description of the strong decay width of non-strange
baryon resonances.

\begin{figure}[ht]
  \begin{center}
         \includegraphics[width=0.40\textwidth]{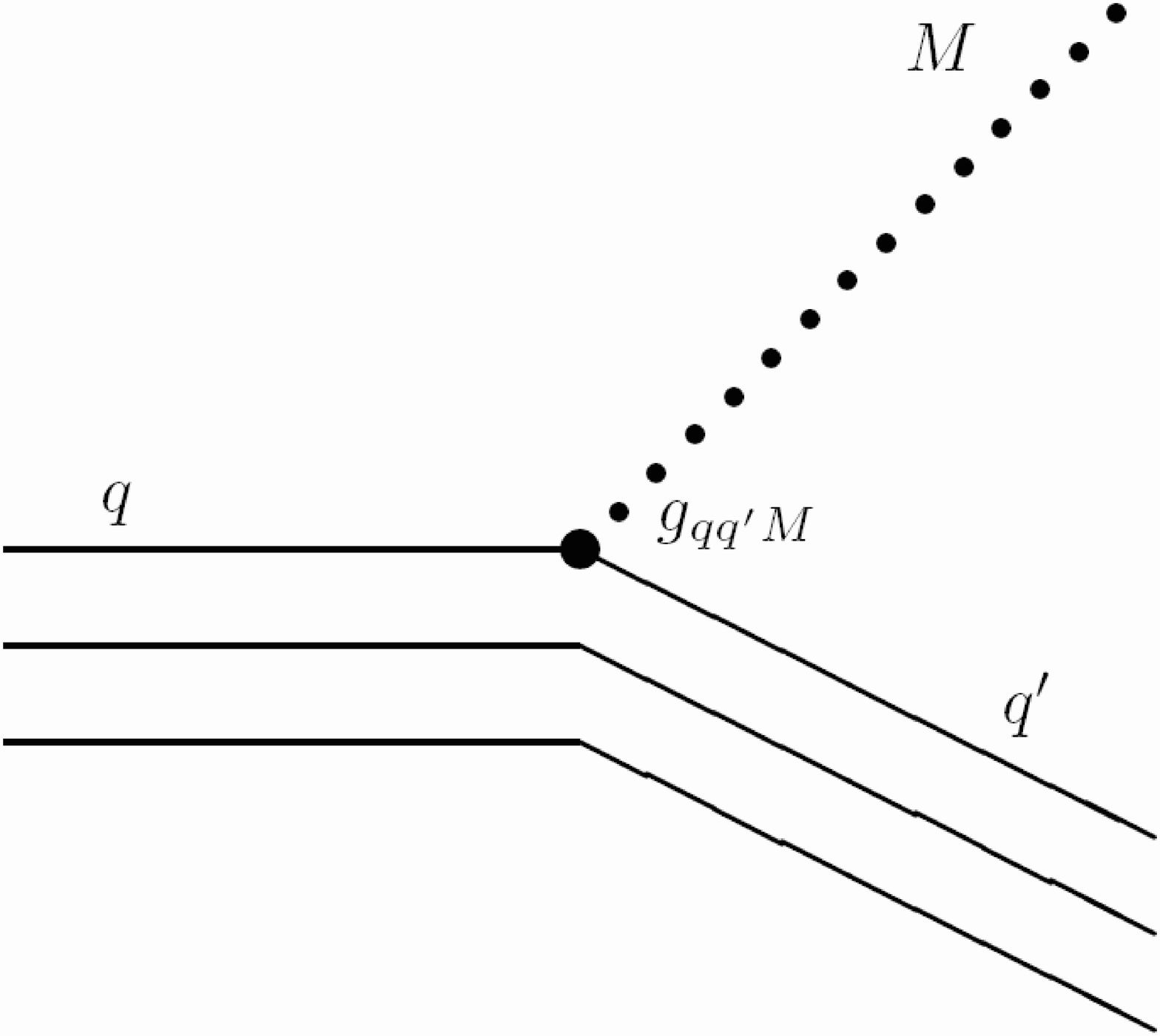}
  {\caption{\label{fig:DecayM2}$B^*\rightarrow B~M$ decay in the point-like meson emission model. }}
  \end{center}
\end{figure}

As well known, the theory of a baryon strongly decay into hadrons is
far away from establishment. In this section, we use a simple model,
called the point-like meson emission model, to estimate the decay
width of the $N^*\to NM$ process. In this model, the baryon has its
own quark structure, and the point-like meson is emitted from one of
the quarks in baryon (see Fig. \ref{fig:DecayM2}).

The amplitude responsible for the emission of a pseudoscalar meson
is usually assumed as
\begin{eqnarray}
<NM\mid H_{s}\mid N^{*}>&=&\sum_{i=1}^{3}<N e^{-i\mathbf{k}\cdot
{\mathbf r}_{i}}
\mid(x\mathbf{k}\cdot\bm{\sigma}_{i}+y\mathbf{p}_{i}\cdot\bm{\sigma}_{i})X_{i,M}\mid
N^{*}> \nonumber \\
&=&3<N e^{-i\mathbf{k}\cdot {\mathbf r}_1}
\mid(x\mathbf{k}\cdot\bm{\sigma}_1+y\mathbf{p}_1\cdot\bm{\sigma}_1)X_{1,M}\mid
N^{*}>, \label{eq:amplitude}
\end{eqnarray}
where the factor 3 is due to the symmetry of the wave function of
three identical quarks, $\mathbf{k}$ is the momentum of the emitted
meson, $\mathbf{r}_i$, $\mathbf{p}_i$,
$\frac{1}{2}\mathbf{\sigma_i}$ and $X_{i,M}$ are the coordinate, the
momentum, the spin and the flavor matrix, that describe the quark
transition process $q_{i}\rightarrow q^{'}_{i}+M$, of the {\it i}-th
quark, respectively, and $x$ and $y$ are phenomenological constants.
For neutral-pion emission, $X_{M}$ can be written as
\begin{eqnarray}
X_{\pi^{0}}=\lambda_{~3},
\end{eqnarray}
where $\lambda_{~3}$ is the Gell-Mann matrix. The wave function of
the $\mid N^{*}>$ state is obtained by the configuration mixing in
calculating the spectrum in the chiral SU(3) quark model.

In the center of mass frame of the non-strange baryon, the width of
a non-strange baryon decaying into a ground state nucleon and a
$\pi$ meson can be written as
\begin{eqnarray}
\Gamma_{N\pi}=\frac{1}{\pi}\frac{|<f\mid \mathcal{H}_{s}\mid
i>|^{2}}{2J_{R}+1}\frac{kE_{N}}{m_{R}}
<I_{N},I_{3N},I_{\pi},I_{3\pi}|I_{{N}^{*}},I_{3{N}^{*}}>^{-2},
\label{eq:decaywidth}
\end{eqnarray}
where $J$, $I$ and $I_3$ represent the total angular momentum, the
isospin and the third component of isospin, respectively.

In Table \ref{tab:decaywidth}, we show the calculated decay width
$\Gamma_{N\pi}^{1/2}(MeV^{1/2})$ of a four-star or a three-star
non-strange resonance decaying into a nucleon and a pion. For
comparison, the results in \cite{decayisgur} are also listed in the
table \ref{tab:decaywidth}. The parameters $x$ and $y$ are obtained
by a $\chi^2$ fit. From the table, one sees that most of the decay
widths in our model agree with the experimental data and are
comparable to those in Ref. \cite{decayisgur}. The decay widths of
$N(\frac{1}{2}^+,1440)$ and $N(\frac{1}{2}^-,1535)$ deviate from the
data greatly, because the order of the masses of these two
resonances in the model prediction are opposite down.  Moreover, the
corresponding decay widths in two confining mode cases are also very
close because the corresponding wave functions are very close.

\begin{table}
\begin{center}
\caption{\label{tab:decaywidth} Square root of the non-strange
baryon decay widths, $\Gamma_{N\pi}^{1/2}$, in MeV$^1/2$, together
with the parameter $x$ and $y$. Column 3 are the results from
\cite{decayisgur}. Column 3 and 4 correspond to $\Delta$-shape and
Y-shape confining potential and the last column are experimental
data from \cite{PDG}.}
\begin{tabular}{ccccc}
\hline\hline
& Isgur&$\triangle-shape$&Y-shape&Expt.\\
x(fm)&&$0.752\times 10^{-2}$&$0.743\times 10^{-2}$& \\
y(fm)&&$-0.845\times 10^{-4}$&$-0.217\times 10^{-3}$&\\
\hline
\multicolumn{1}{c}{Resonance}&\multicolumn{4}{c}{$\Gamma_{N\pi}^{1/2}(MeV^{1/2})$}\\
\hline
$N\frac{1}{2}^{-}(1535)$&5.3&13.2&13.3&$8.2\pm 2.3$\\
$N\frac{1}{2}^{-}(1650)$&8.7&8.6&8.2&$10.4 \pm 2.0$\\
$N\frac{3}{2}^{-}(1520)$&9.2&11.5&10.9&$8.1 \pm 0.8$\\
$N\frac{3}{2}^{-}(1700)$&3.6&2.9&2.9&$3.2 \pm 1.6$\\
$N\frac{5}{2}^{-}(1675)$&5.5&6.7&6.4 &$8.2 \pm 1.0$\\
$N\frac{1}{2}^{+}(1440)$&6.8&6.6& 6.1&$15.1 \pm 2.8$\\
$N\frac{1}{2}^{+}(1710)$&6.7&3.1&3.2 &$3.9 \pm 2.4$\\
$N\frac{3}{2}^{+}(1720)$&6.5&3.4&3.8 &$4.7 \pm 1.6$\\
$N\frac{5}{2}^{+}(1680)$&7.1&2.4&2.6 &$9.2 \pm 0.7$\\
$\Delta\frac{1}{2}^{-}(1620)$&3.3&5.7&5.7 &$ 6.1\pm 1.2$\\
$\Delta\frac{3}{2}^{-}(1700)$&4.9&6.4&6.9 &$ 6.7\pm 2.2$\\
$\Delta\frac{1}{2}^{+}(1910)$&5.3&5.7&4.9 &$7.5 \pm 1.8$\\
$\Delta\frac{3}{2}^{+}(1232)$&10.2&10.3&10.7 &$10.9 \pm 0.2$\\
$\Delta\frac{5}{2}^{+}(1905)$&4.0&5.5&5.7 &$5.9 \pm 2.2$\\
$\Delta\frac{7}{2}^{+}(1950)$&7.5&10.6&10.4 &$10.6 \pm 0.9$\\
\hline\hline
\end{tabular}
\end{center}
\end{table}

\section{Conclusions}

The baryon spectra with both Y-mode and $\Delta$-mode confining
potentials in $N\leq 2$ bands are studied in the chiral $SU(3)$
quark model. The results show that, no matter which type of
confining potential, say the $\Delta$-mode or the Y-mode or even the
mixed mode, is employed, the experimental baryon spectra can be
well-explained. The resultant baryon spectra with the Y-mode and
$\Delta$-mode confining potentials are very close to each other. For
most of the states, the corresponding mass difference with different
confining modes is less than 20 MeV. Moreover, the $\Delta$-mode
confinement is more effective in the short and medium distances and
the Y-mode confinement provides more contributions in the long
distance. It is also shown that the effect of the different
confining mode does not distinctly show up in the spectrum. The
strong decay widths of the non-strange baryon resonances are
calculated in the point-like meson emission model with the
wave-functions obtained in the baryon spectrum calculation. The
resultant decay widths are generally in agreement with the
experimental data. However, this mode is too simple. The detailed
decay information of baryons should be extracted from the
investigation in the implicit quark-gluon degrees of freedom.

This work is partially supported by the National Natural Science
Foundation of China under grant Nos. 10475089, 10435080, 10375090
and CAS grant No. KJCX3-SYW-N2.

\end{document}